\documentclass[12pt,amsfonts,enumerate,amscd,reqno]{amsart}
\usepackage{enumerate}
\usepackage{amsthm}
\usepackage{graphicx}

\numberwithin{equation}{section} 

\newtheorem{thm}{Theorem}[section]
\newtheorem{cor}[thm]{Corollary}
\newtheorem{lem}[thm]{Lemma}

\newcommand{\lemm}[1]{Lemma \ref{#1}}
\newcommand{\theor}[1]{Theorem \ref{#1}}

\newcommand{\sect}[1]{Section \ref{#1}}
\newcommand{\subsect}[1]{Subsection \ref{#1}}
\newcommand{\eq}[1]{(\ref{#1})}

\newcommand{\bbe}{\begin{equation}}
\newcommand{\lan}{\langle}
\newcommand{\ran}{\rangle}

\newcommand{\Tr}{{\rm Tr}}

\newcommand{\R}{{\bf R}}
\newcommand{\C}{{\bf C}}

\newcommand{\cB}{{\mathcal B}}

\newcommand{\cF}{{\mathcal F}}

\newcommand{\LL}{{\mathcal L}}

\newcommand{\pdt}{{\partial_t}}

\newcommand{\rn}{\R^n}
\newcommand{\hu}{\hat u}

\newcommand{\hd}{\hat d}

\newcommand{\hF}{\hat F}

\newcommand{\eps}{\epsilon}
\newcommand{\de}{\delta}
\newcommand{\al}{\alpha}
\newcommand{\be}{\beta}

  \newcommand{\sg}{\sigma}

\newcommand{\ee}{\end{equation}}

\newcommand{\dd}{\partial}
\newcommand{\la}{\lambda}

\newcommand{\ka}{\kappa}

\newcommand{\om}{\omega}

\newcommand{\ga}{\gamma}
\newcommand{\gap}{\gamma_+}
\newcommand{\gam}{\gamma_-}

\newcommand{\tr}{\tilde r}

\newcommand{\tka}{{\tilde \ka}}
\newcommand{\tbe}{{\tilde \be}}
\newcommand{\td}{{\tilde d}}

\newcommand{\supp}{{\rm supp \ }}

\begin{document}

\title[Consistency conditions for ATSM]{Consistency conditions for affine term structure models}
\author[S.~Levendorski\v{i}]{Sergei Levendorski\v{i}}\thanks{
The author is grateful to Darrel Duffie for the indication that
the use of the Feyman-Kac theorem in ATSM is not justified even
for many diffusion processes, and illuminating discussions. The
author is thankful to the participants of a seminar at the Finance
Department of the University of Texas at Austin for useful
remarks, and to the referee of the paper for useful suggestions.}
\maketitle
 \centerline{Department of Economics, }

\centerline{University of Texas at Austin}

\begin{abstract}
ATSM  are widely applied for pricing of bonds and interest rate
derivatives but the consistency of  ATSM when the short rate, $r$,
is unbounded from below remains essentially an open question.
First, the standard approach to ATSM uses the Feynman-Kac theorem
which is easily applicable only when  $r$ is bounded from below.
Second, if the tuple of state variables belongs to the region
where $r$ is positive, the bond price should decrease in any state
variable for which the corresponding coefficient in the formula
for $r$ is positive; the bond price should also decrease as the
time to maturity increases. In the paper, sufficient conditions
for the application of the Feynman-Kac formula, and monotonicity
of the bond price are derived, for wide classes of affine term
structure models in the pure diffusion case. Necessary conditions
for the monotonicity are obtained as well. The results can be
generalized for jump-diffusion processes.
\end{abstract}


\newpage
\section{Introduction}

Consider a financial market under several sources of uncertainty
represented by a multi-variate Markov process $X$. The price of an
interest rate derivative of the European type, maturing at date
$T$, with the terminal pay-off $g(X(T))$, can be expressed as
\begin{equation}\label{defcc}
f(X(t), t)= E_t\left[\exp\left(-\int_t^T
r(X(s))ds\right)g(X(T))\right].
\end{equation}
Starting with one-factor diffusion models Vasicek (1977) and Cox,
Ingersoll and Ross (1985), one of the popular approaches has been
 to model $X$ as the solution to the
stochastic differential equation
\begin{equation}\label{st1}
dX_j(t)=b_j(X(t), t)dt + \sum_{k=1}^n\be_{jk}\sqrt{S_j(X(t),
t)}dW_j,
\end{equation}
$j=1,\ldots, n$, where $b_j$ and $S_j$ are affine functions of
$X(t)$, and $dW$ is the increment of the standard $n$-dimensional
Brownian motion; $r$ is modelled as an affine function of the
state variable:
\begin{equation}\label{affr}
r(X(s))=\lan d, X(s)\ran + d_0,
\end{equation}
where $d\in\rn$ is a constant vector, and $d_0\in\R$ is a scalar;
$\lan\cdot,\cdot\ran$ denotes the standard inner product in $\rn$.
 When $r$ is given by \eq{affr},
the Feynman-Kac formula and the Fourier transform can be used to
reduce the calculation of $f(X(t), t)$ to the solution of a
parabolic equation, and then to solution of a system of ODE
(Riccati equations) depending on a parameter, with the subsequent
integration w.r.t. this parameter. This idea is due to Heston
(1993) who applied it to pricing of bond and currency options.
Heston's approach was generalized by Duffie and Kan (1996), who
coined the term {\em Affine Term Structure Models} (ATSM). For the
classification of ATSM under diffusion processes, see Dai and
Singleton (2000), and for the extension of ATSM to some
jump-diffusion processes and extensive bibliography on different
families of ATSM for both pure jump and jump-diffusion cases, see
e.g. Duffie, Pan and Singleton (2000) and Chacko and Das (2002).
Notice that the presence of jumps imposes additional restrictions
on the parameters of the model. For instance, in one-dimensional
case,
 one must ensure
that  jumps cannot move $X(t)$ in the region where the volatility
coefficient becomes negative. Thus, either the volatility is
independent of the state variable, or an appropriate restriction
on the direction of jumps must be imposed. For very general
classes of affine Markov models with jumps, under conditions which
ensure the non-negativity of $r$, see Duffie, Filipovi\'c and
Schachermayer (2002).

The consistency of ATSM in cases when $r$ may be unbounded from
below remains essentially an open question. The main stress in the
classification paper Dai and Singleton (2000) is on the
overdeterminacy of many ATSM models; however, for wide regions in
the parameter's space, standard ATSM models may be inconsistent,
and the following issues must be addressed. First, the standard
approach is based on the Feynman-Kac formula but the general
Feynman-Kac theorem is formulated for bounded (and sufficiently
regular, say, continuous) $r$ and sufficiently regular $g$; an
extension to the case of $r$ which are bounded from below is
straightforward. Thus, the first step of the solution of an ATSM,
namely, the reduction to the backward parabolic problem
\begin{eqnarray}\label{fc1}
(\pdt +L -r)f(x, t)&=&0,\quad t<T;\\\label{bc1} f(x, T)&=&g(x),
\end{eqnarray}
where $L$ is the infinitesimal generator of $X$, cannot be easily
deduced from a general Feyman-Kac theorem unless the affine $r$
depends on $X_j$ of the CIR-type only\footnote{We say that a
factor $X_j$ in an ATSM is of the CIR-type iff it assumes values
in $\R_+$}.
  If some of
$X_j$ may assume  arbitrary (real) values, and the corresponding
coefficients $d_j$ in \eq{affr} are non-zero, then $r$ is
unbounded from below, and to the best of our knowledge, no
universal statement exists which guarantees that the solution to
the problem \eq{fc1}-\eq{bc1}, call it $f_0(g, r; x, t)$,
coincides with $f(g, r; x, t)$ given by the stochastic expression
\eq{defcc}.
 Second, it is natural to assume that in the fully consistent
model, the solution to the bond pricing problem must be a
decreasing function of any state variable for which the
corresponding coefficients in the formula for $r$ are positive;
the solution must also decrease as the time to maturity increases,
if the tuple of state variables belongs to the region where $r$ is
positive.

In the classification of Dai and Singleton (2000), a model is said
to belong to family $A_m(n)$ if the number of the factors of the
CIR-type is $m$. The two opposite cases, $m=n$ and $m=0$, are
especially simple. If $m=n$, then the justification of the use of
the Feynman-Kac theorem is a special case of the general result,
and if in addition, $d_0\ge 0$, then the monotonicity of the bond
price is evident from \eq{defcc}. If $m=0$, then  $X$ is the
Ornstein-Uhlenbeck process on $\rn$,  for which an explicit
formula for the characteristic function $E^x\left[e^{i\langle \xi,
X(t)\rangle}\right]$ is available (see equation (17.4) in Sato
(1999)\footnote{The author is grateful to Darrell Duffie for the
reference}). The formula holds even for a wider class of
 processes of Ornstein-Uhlenbeck type, which are driven by L\'evy
processes, and it is equivalent to the statement that the
stochastic expression \eq{defcc} with $g\equiv 1$  is equal to the
solution to the Riccati equations for the bond price, that is,
$f_0(1, r; x, t)=f(1, r; x, t)$. However, there is no reason to
expect that the bond price is monotone in this case for all
parameters' values. For $1\le m \le n-1$, both the reduction to
the Riccati equations and the monotonicity conditions have not
been studied.

Let $\tau=T-t$ be the time to maturity, and
\begin{equation}\label{deff0}
P(x, \tau)=\exp\left[\sum_{j=1}^n B_j(\tau)x_j+C(\tau)\right]
\end{equation}
be the price of the bond, which is obtained in an ATSM model by
the formal reduction to the Riccati equations.  First, we consider
the Vasicek model and its generalization, namely, family $A_0(n)$,
then a simple two-factor $A_1(2)$ model, next more general
$A_1(n)$ model, and finally the family $A_2(3)$ (other families
$A_m(n)$ can be studied similarly), and derive, in terms of
parameters of the model,
\begin{enumerate}[(I)]
\item
 simple necessary conditions for the decay of $P(x, \tau)$:
\begin{equation}\label{bonddecay}
P'_\tau(x, \tau)<0,\quad \forall\ x\quad {\rm s.t.}\ r(x)\ge
0,\quad {\rm and}\ \forall\ \tau>0;
\end{equation}
in some cases, we also show that these conditions imply the
boundedness of the bond price: in the region $\{x\ |\ x_j>0:
d_j>0\}$,
\begin{equation}\label{bondbound}
P(x, \tau)<C,\quad \forall\ \tau>0;
\end{equation}
\item sufficient conditions for the decay of $B_j(\tau)$; we do not know
how wide is the gap between these conditions and the (unknown to
us) necessary and sufficient conditions;
\item
sufficient conditions under which  the reduction to the system of
the Riccati equations can be justified. For $A_1(2)$ family, and
in many other cases, these condition are weaker than the necessary
conditions in (I).
\end{enumerate}
{\em Remark 1.1.} a) As it was mentioned above, for
$A_0(n)$-model, the reduction to the system of Riccati equations
is known, and it is valid without additional conditions on
parameters of the model.

 b)
Necessary and sufficient conditions for \eq{bonddecay} in a
neighborhood of $\tau=+\infty$, and in a vicinity of 0, are easier
to derive, and under these conditions, a ``numerical proof" of the
monotonicity of the bond price on a large finite interval can be
used to show that for given parameters' values, the model is
consistent.

c) As our study shows, for  family $A_1(n)$, the monotonicity of
$P(x, \tau)$ w.r.t. $\tau$ is the main consistency problem for
ATSM (and the only consistency problem for family $A_0(n)$). On
the other hand, should one use the model for a fixed (and
sufficiently small) time to maturity then the model can be
consistent on this time interval; and it is possible to derive
sufficient conditions for \eq{bonddecay} to be valid on a
sufficiently narrow interval $(0, \tau_0)$, where $\tau_0>0$
depends on parameters of the model.

d) When it is necessary to consider more general contingent
claims, a sufficient condition for (III), in terms of the rate of
growth of the pay-off at infinity, can be derived relatively
easily, and the same is true of a necessary condition for the
natural analog of \eq{bonddecay} and sufficient condition for
\eq{bondbound}. The sufficient conditions for the monotonicity
will be more difficult to derive.

e) It is plausible that in some empirical studies, the fitted ATSM
is inconsistent in the sense that the monotonicity condition
fails. Hence, if the model is fitted for some time to maturity,
and used later for a larger time, then it may produce
non-monotonic bond prices.

f) Similar consistency problems exist for interest rate derivative
products, and an unnatural behavior of the price of a derivative
product can be easily overlooked if one fits the parameters of the
model by using the data on bond prices, and then uses the
calibrated model to calculate prices of interest rate derivatives.
It might be possible to construct an arbitrage strategy against a
counterparty who uses an inconsistent model.

 One
may argue that the consistency analysis should be conducted when
not only parameters of the model are fixed but values of
(unobservable) factors as well: if for a chosen set of the
parameters and factors the bond price is a decaying function of
$\tau$, then the model is reasonable. However,  it is not clear
how to obtain a general result in this set-up, and moreover, it
does not seem to be right to put the factors on the equal footing
with the parameters of the model, especially in cases when the
factors can be interpreted as the short rate, its volatility
and/or central tendency.

We assume that the model should be consistent for all positive
values of factors, the interest rate depends on (we consider the
case when in \eq{affr}, the coefficients  $d_j, j=1,\ldots, n,$
are non-negative), which makes the model less flexible than it is
assumed in Dai and Singleton (2000).  For the factors of the
CIR-type, our restriction on $d_j$'s is natural, and it is without
loss of generality as far as the factors assuming values in $\R$
are concerned: if one of these $d_j$ is negative, one can make it
positive by using $-X_j$ instead of $X_j$. On the other hand, the
model specifications are underdetermined in the sense of Dai and
Singleton (2000) because this allows us to formulate necessary
conditions and sufficient ones in  more symmetric and natural
forms.

The rest of the paper is organized as follows. In \sect{a0n}, we
formulate and prove the results for the Vasicek model and its
generalization, $A_0(n)$- model. Although the justification of the
use of the Feyman-Kac theorem can be made by appealing to formula
(17.4) in Sato (1999),  we give an independent proof of the
reduction.  The main idea of the proof is the conjugation with an
appropriate exponent, which allows us to reduce to the case of an
interest rate bounded from below. The same idea is used in the
proofs for $A_m(n)$-models, $m\ge 1$, however the realization
becomes much more involved, if we want to avoid too stringent
conditions.

In \sect{main}, we formulate the main results for the case $m\ge
1$, and prove theorems about properties of the formal solution
\eq{deff0} of the ATSM model.
In \sect{genjust}, we describe the general scheme of justification
of the reduction to the backward parabolic equation; in other
words, the scheme of the proof of the Feynman-Kac formula for an
affine $r$ and the class of processes used in $A_m(n)$ models,
$m\ge 1$. The main ingredient of the proof is the representation
theorem for analytic semigroups. The proof of the reduction for
$A_1(n)$ model and $A_2(3)$ model is given in \sect{reductiona1n}
(more general $A_m(n)$-models can be studied similarly).  The most
technical part of the proof, namely,
 the proof of  the existence and uniqueness theorem for degenerate
elliptic operators with parameter is delegated to the appendix.
The proof of the latter theorem uses a general approach to
problems of this sort described in detail in the review paper
Levendorski\v{i} and Paneyakh (1990) and monograph
Levendorski\v{i} (1993). This approach is applicable not to
differential operators only but to integro-differential operators
(another name: pseudo-differential operators) as well, which
allows one to justify the use of the Feynman-Kac formula for
jump-diffusion processes. This more general case will be treated
in a separate publication.

\section{Family $A_0(n)$}\label{a0n}
\subsection{Monotonicity conditions}
The interest rate is given by \eq{affr} with non-negative $d_j,
j\ge 1,$ and positive $d_n$, and the dynamics of $X(t)$ is given
by SDE
\begin{equation}\label{sde0n}
dX(t)=(\theta-\ka X(t))dt + \Sigma dW(t),
\end{equation}
where $W(t)$ is the standard Brownian motion in $\rn$,
$\ka=[\ka_{jl}]$ is a low-diagonal matrix with positive diagonal
elements, $\Sigma$ is a positive-definite matrix, and $\theta$ is
a constant vector. The bond price is given by \eq{deff0} with
\begin{equation}\label{solB0n}
B(\tau)=-(1-\exp(-\tau \ka^T))\ga,
\end{equation}
where $\ga=(\ka^T)^{-1}d$, and
\begin{equation}\label{solC0n}
C(\tau)=\int_0^\tau \left(-d_0+\theta^T B(s)+\frac{1}{2}\Tr\left(
\Sigma^T[B_j(s)B_l(s)]_{j,l=1}^n\Sigma\right)\right)ds,
\end{equation}
and necessary and sufficient conditions for the decay of the bond
price are (relatively) easy to establish. These conditions are
especially simple in the case $n=1$ (the Vasicek model).
\begin{thm}\label{Vas}
Let $n=1$, and let the dynamics of the short rate be given by
\[
dr(t)=(\theta-\ka r(t))dt + \sg dW(t),
\]
where $\theta, \ka$ and $\sg$ are positive constants.

Then a) if for some $r>0$, the bond price $P(r, \tau)$ is a
non-increasing function of $\tau$, then
\begin{equation}\label{necsufvac}
\frac{\sg^2}{2}\le \ka \theta;
\end{equation}
b) if \eq{necsufvac} holds, then for any $r>0$, the bond price
$P(r, \tau)$ is a decreasing function in $\tau$.
\end{thm}
\begin{proof}
Clearly, $B(\tau)=-\ka^{-1}(1-e^{-\ka\tau})$ is decreasing on $[0,
+\infty)$ from 0 to $-\ka^{-1}$, therefore if
\[
C'(\tau)=\theta B(\tau)+\frac{\sg^2}{2}B(\tau)^2
\]
is non-positive on $(0, +\infty)$, then \eq{necsufvac} holds, and
if  \eq{necsufvac} holds, then $C'(\tau)$ is negative on $(0,
+\infty)$.
\end{proof}
Condition \eq{necsufvac} means that for a given central tendency
and coefficient of mean reversion, the volatility may not be too
large, and the interpretation is clear: if volatility is large,
then a trajectory of the process spends significant amount of time
in a region of the state space where the short rate is negative,
which leads to the artificial increase of the bond price.

 In the case $n\ge 2$, it is convenient to study the monotonicity
of $B_j$ step by step, by using the low-diagonal structure of
$\ka$, instead of appealing to the explicit general formula
\eq{solB0n}. Due to  positivity of the diagonal elements of matrix
$\ka$ and the assumption $d_n>0$, the $B_n$ is decreasing from 0
to $-d_n/\ka_{nn}$. Hence, if \[ B_{n-1}(\tau)=\int_0^\tau
e^{-\ka_{n-1, n-1}(\tau-s)}(-\ka_{n, n-1}B_n(s)-d_{n-1})ds \] is
non-increasing on $(0, +\infty)$, then it must be that
\begin{equation}\label{necsuf02}
d_{n-1}\ka_{nn}-d_n \ka_{n, n-1}\ge 0, \end{equation} and if
\eq{necsuf02} holds, then $B_{n-1}$ decreases. For $B_j, j\le
n-2,$ the necessary conditions for the monotonicity are not that
simple but the induction and the same consideration as above show
that if the off-diagonal entries of the matrix $\ka$ are
non-positive then all $B_j$'s are decreasing functions. From
\eq{solC0n}, it is evident that then the bond price is a
decreasing function in $\tau$ if and only if the function
\[
\R_+\ni s\mapsto F(s):=-d_0+\langle\theta, B(s)\rangle +
\frac{1}{2}\Tr\left(
\Sigma^T[B_j(s)B_l(s)]_{j,l=1}^n\Sigma\right)\in \R
\]
is non-positive. Clearly,  necessary and sufficient conditions for
non-positivity of $F$ in terms of the parameters of the model
cannot be simple, however, relatively simple necessary conditions
and sufficient ones (the latter more stringent than the former)
are easy to formulate. From \eq{solB0n}, we find that
$B(+\infty)=-\ga:=-(\ka^T)^{-1}d$, and therefore for any
$j=1,\ldots, n$, and $s>0$, we have $B_j(s)\in (-\ga_j, 0)$.
Introduce the quadratic polynomial \[ Q(y)=-d_0+\langle\theta,
y\rangle +\frac{1}{2}\Tr\left(
\Sigma^T[y_jy_l]_{j,l=1}^n\Sigma\right). \] If $Q(-\ga)>0$, then
for sufficiently large $s$, $F(s)>0$, and therefore $C$ grows in a
neighborhood of $+\infty$; this gives a necessary condition
 \begin{equation}\label{nec1n}
 Q(-\ga)\le 0.
\end{equation}
Since the quadratic form in the definition of $Q(y)$ is
positive-definite, $Q$ is non-positive on $U^\ga:=[-\ga_1,
0]\times\cdots \times [-\ga_n, 0]$ (which implies that the
integrand in \eq{solC0n} is negative a.e., and $C$ decreases) if
and only if it is non-positive at each vertex of $U^\ga$, that is,
\begin{equation}\label{suf0n}
Q(y)\le 0\quad \forall\ y\in \{-\ga_1, 0\}\times\cdots
\times\{-\ga_n, 0\}.
\end{equation} We have obtained
\begin{thm}\label{necsuf0n}
a) Let $n=2$, and let $B_1$ be non-increasing.

 Then \eq{necsuf02}
holds.

b) Let $n\ge 2$, and let all $B_j$ be non-increasing. Then
\eq{nec1n} holds.

 c) Let $n\ge 2$, let the off-diagonal entries of $\ka$ be
non-positive, and let \eq{suf0n} hold.

 Then the bond price is a
decreasing function of $(x, \tau)$ in the region $x>0, \tau>0$.
\end{thm}

\subsection{Justification of the Feyman-Kac formula}
We assume that $\Sigma$ is non-degenerate.
\begin{thm}\label{fka0n}
Let $g$ be a continuous function, which does not grow too rapidly
at the infinity
\begin{equation}\label{infdec}
\ln (1+|g(x)|)=o(||x||^2), \quad {\rm as}\ x\to+\infty.
\end{equation}
Then the  expressions \eq{defcc}, \eq{st1} and \eq{affr} are the
unique solution to the problem \eq{fc1}-\eq{bc1} in the class of
continuous functions, which admit the bound \eq{infdec} uniformly
in $t\in [0, T]$.
\end{thm}
As it will be seen from the proof,  condition on $g$ can be
relaxed, and the solution is unique in a wider class of functions.
\begin{proof}
 By making an affine transformation of the factors, we may assume
that there exist $c_0>0$ such that
\begin{equation}\label{kac0}
\langle\ka x, x\rangle\ge c_0||x||^2.
\end{equation}
Denote by $f_0(g, r,; x, t)$ a solution to the problem
\eq{fc1}-\eq{bc1}.
\begin{lem}\label{a0nconv}
a) Let $r$ and $g$ be continuous, let $g$ satisfy \eq{infdec} and
$r$ satisfy
 \begin{equation}\label{decr}
 r(x)=o(||x||^2), \quad {\rm as}\ x\to \infty.
 \end{equation}
  Then a solution to the problem \eq{fc1}-\eq{bc1} in
the class of continuous functions $f(x, t)$, which admit the bound
\eq{infdec} uniformly in $t\in [0, T]$, exists and it is unique.

b) Let $r_1, r_2, \ldots, r_N,\ldots$, be a sequence of continuous
functions which satisfies \eq{decr} uniformly in $N$, and
converges pointwise to a function $r$. Then
\begin{equation}\label{conv0n}
f_0(g, r_N; x, t)\to f_0(g, r; x, t)\quad {\rm as}\quad
N\to+\infty,
\end{equation}
pointwise.
\end{lem}
\begin{proof}
The key element of our approach  is the conjugation with an
appropriate exponential function; in the case of $A_0(n)$, this
function is especially simple. Take a small $\eps>0$, and set
\begin{eqnarray*}
f_\eps(g, r; x, t)&=&\exp(-\eps||x||^2)f_0(g, r; x, t),\\
g_\eps(x)&=&\exp(-\eps||x||^2)g(x).
\end{eqnarray*}
Similarly, define $f_\eps(g, r_N; x, t)$. Due to \eq{infdec},
$f_\eps(g, r; x, t)$, $f_\eps(g, r_N; x, t)$ and $g_\eps(x)$
vanish as $x\to \infty$, uniformly in $N$ and $t\in [0, T]$.

a) Insert $f_0(g, r; x, t)=\exp(\eps||x||^2)f_\eps(g, r; x, t)$
into \eq{fc1}-\eq{bc1}, and multiply by $\exp(-\eps||x||^2)$. The
result is a problem of the same form
\begin{eqnarray}\label{fc10}
(\pdt +L_\eps -r_\eps)f_\eps(g, r, x, t)&=&0,\quad
t<T;\\\label{bc10} f_\eps(g, r, x, T)&=&g_\eps(x),
\end{eqnarray}
where $r_\eps$ is a function, and $L_\eps$ is a differential
operator without the first order term, which are obtained from
\begin{equation*}\label{conjn}
\exp(-\eps||x||^2)(L-r)\exp(\eps||x||^2)=L_\eps-r_\eps.
\end{equation*}
It is easily seen that the coefficients of $L_\eps$ at derivatives
of order 2 are the same as the ones of $L$, and the other
coefficients of $L_\eps$ tend to the corresponding coefficients of
$L$ as $\eps\to 0$. Hence, if $\eps>0$ is sufficiently small, then
$L_\eps$ is the infinitesial generator of an Ornstein-Uhlenbeck
process, call it $X^\eps$. Fix such an $\eps$. Further,
$r_\eps=r+\tr_\eps$, where
\[
\tr_\eps(x)=2\eps\langle \ka x, x\rangle +
\eps(\Tr(\Sigma^T\Sigma)-2\langle\theta, x\rangle),
\]
and in view of \eq{decr} and \eq{kac0}, there exist positive $c,
C$ such that
\begin{equation}\label{reps}
c||x||^2-C\le r_\eps(x)\le c||x||^2+C,\quad \forall \ x\in\rn.
\end{equation}
Since $r_\eps$ is bounded from below, the solution to problem
\eq{fc10}-\eq{bc10}  exists in the class of continuous functions
decaying as $x\to\infty$, and it is unique (in fact, the solution
is unique in the class of functions which grow not faster than an
exponential function).
 Since $f_0(g, r; x, t)=\exp(\eps||x||^2)f_\eps(g, r; x, t)$, part
 a) has been proved.

 b) The argument above is applicable with $r_N$ instead of $r$,
 and $r_{N,\eps}=r_N+\tr_\eps$ admits bound \eq{reps} uniformly in
 $N$. It follows that $f_\eps(g, r_N, x, t)$ is given by  the Feynman-Kac formula
\begin{equation}\label{fcsol}
f_\eps(g, r_N, x, t)=E_t\left[\exp(-\int_t^T r_{N,\eps}(X^\eps(s))
ds)g_\eps(X^\eps(T))\right].
 \end{equation}
 Since
$r_{N,\eps}(x)\to r_\eps(x)$, as $N\to +\infty$, point-wise, we
use the Dominant Convergence Theorem, pass to the limit in
\eq{fcsol}, and obtain that $f_\eps(g, r_N, x, t)\to f_\eps(g, r,
x, t)$, point-wise. It follows that $f_0(g, r_N, x, t)\to f_0(g,
r, x, t)$, point-wise.
\end{proof}
Now we give the proof of \theor{fka0n}. Denote $f(g, r; x, t)$ as
the stochastic expression  \eq{defcc}. Fix $N\in\R$, and for the
affine $r$, set $r_N(x)=\max\{r(x), -N\}$. Then $r_N$ is bounded
from below, hence both $f(g, r_N; x, t)$ and $f_0(g, r_N; x, t)$
exist, and $f(g, r_N; x, t)=f_0(g, r_N; x, t)$ for all $x$ and
$t<T$. Since $g$, $r_N$ and $r$ satisfy the conditions of
\lemm{a0nconv}, \eq{conv0n} holds.
 By the Monotone Convergence Theorem,
$ f(g, r_N; x, t)\to f(g, r; x, t)\quad {\rm as}\ N\to +\infty,$
point-wise, and we conclude that $f(g, r; x, t)=f_0(g, r; x, t)$
for all $x$ and $t\le T$.
\end{proof}

\section{Families $A_m(n)$, $1\le m\le n-1$.}\label{main}

\subsection{Family $A_1(2)$}\label{maina12} The state space is $\R_+\times \R$,
the $r$ is given by \eq{affr} with $d_1\ge 0, d_2>0$, and the
infinitesimal generator of the process is of the form
\begin{equation}\label{gena12}
L=(\theta_1-\ka_{11}x_1)\dd_1+(\theta_2-\ka_{21}x_1-\ka_{22}x_2)\dd_2+\frac{1}{2}
x_1\dd_1^2+\frac{\al+\be x_1}{2}\dd_2^2,
\end{equation}
where $\ka_{11}, \ka_{22}, \theta_1,  \al, \be$ are positive.
 Without loss of generality, $\theta_2=0$.

  Assume that in the case
of the bond, the use of the Feynman-Kac theorem has been
justified. Then $P(x, \tau)=f_0(1, r; x, t)$, the solution to
\eq{fc1}-\eq{bc1} with $g(x)\equiv 1$, can be found in the form
\eq{deff0}. By substituting \eq{deff0} into \eq{fc1}-\eq{bc1}, we
obtain the system of Riccati equations on $(0, T)$:
\begin{eqnarray}\label{B1}
B_1'&=&-\ka_{11} B_1+\frac{1}{2} B_1^2-\ka_{21}
B_2+\frac{\be}{2}B_2^2-d_1,\\\label{B2} B_2'&=&-\ka_{22}
B_2-d_2,\\\label{C} C'&=& -d_0+\theta_1 B_1+\frac{\al}{2}B_2^2,
\end{eqnarray}
subject to the boundary conditions
\begin{equation}\label{Ricb}
B_1(0)=0;\quad B_2(0)=0;\quad C(0)=0.
\end{equation}
We solve \eq{B2} subject to  $B_2(0)=0$:
\begin{equation}\label{solB2}
B_2(\tau)=-\ga\left(1-e^{-\ka_{22}\tau}\right),
\end{equation}
where $\ga:=\ka_{22}^{-1}d_2>0$. $B_2$ decreases from 0 to $-\ga$,
since
    $\ka_{22}>0$:
 \begin{equation}\label{B2lim}
  \lim_{\tau\to +\infty} B_2(\tau)=-\ga<0.
  \end{equation}
 If $B_1(\tau_0)>0$ for some $\tau_0>0$, then not only $P(x, \tau)$
  fails to be a decaying function of  $\tau$; for this $\tau_0$, $P(x, \tau)$
  is an increasing function in $x_1$. The following theorem
  provides necessary conditions which exclude such
  a
  strange behavior of the bond price (the higher the spot short rate,
  the higher the price of the bond), and  sufficient conditions for the
  negativity of $B_1(\tau)$ for $\tau>0$.
  \begin{thm}\label{thm:necess1}
 a) If $B_1$ is non-increasing on $[0, +\infty)$, then the following
 two conditions hold:
 \begin{equation}\label{kad}
 d_1>0\quad {\rm or}\quad \ka_{21}<0;
 \end{equation}
\begin{equation}\label{suff0}
d_1-\ka_{21}\ga- \frac{\be}{2}\ga^2\ge 0.
 \end{equation}
b) If \eq{kad}-\eq{suff0} hold, then
 \begin{equation}\label{B1neg}
 B_1(\tau)<0,\quad \forall\ \tau>0.
 \end{equation}
 \end{thm}
\begin{proof}
 a) Suppose that $d_1=\ka_{21}=0$ but $B_1(\tau)$ is non-positive
in a  right neighborhood of 0. From \eq{solB2}, the RHS in \eq{B1}
is positive in this neighborhood, hence $B_1$ is increasing in
this neighborhood from 0. Hence, $B_1$ is positive there;
contradiction. Thus, \eq{kad} holds.

 Denote by $\hd$ the LHS of \eq{suff0}.
 In view of \eq{B2lim},
the RHS of \eq{B1} admits the representation
\[
-\ka_{11}B_1(\tau)+\frac{1}{2}B_1(\tau)^2-\hd +o(1),\quad {\rm
as}\ \tau\to +\infty.
\]
Hence, if $\hd$ is negative, and $B_1$ stays negative, then
eventually, the RHS of \eq{B1} will exceed a positive constant,
and $B_1$ will grow as a linear function of $\tau$, contradiction.

b) From \eq{kad} and \eq{solB2}, we see that $B_1$ is negative in
a right neighborhood of 0.  Consider the quadratic
 polynomial $Q(y)=-d_1-\ka_{21}y+\frac{\be}{2}y^2$. Under condition \eq{kad},
 it is negative in a small left neighborhood of 0, therefore if \eq{suff0} holds,
 it is negative on $(-\ga, 0)$. Hence,
 \[
 -d_1-\ka_{21}B_2(\tau)+\frac{\be}{2}B_2(\tau)^2<0, \quad\ \forall\
 \tau>0.
 \]
 We conclude that if $B_1(\tau)$ tries to approach zero
from below as $\tau$ increases, then the RHS in \eq{B1} becomes
negative, and pushes the trajectory $\tau\mapsto B_1(\tau)$ down.
Hence, $B_1(\tau)$ remains negative.
\end{proof}
 The sufficient conditions for the monotonicity are
 stronger than \eq{kad}--\eq{suff0}; in fact, additional necessary
 conditions can be derived. We are satisfied with
 \eq{kad}--\eq{suff0} for the time being because these conditions are more
 restrictive than the sufficient conditions for the use of the
 Feynman-Kac formula in \theor{thm:reduction1} below. The following
 theorem provides sufficient conditions for the decay of $B_1$ and $C$.
 \begin{thm}\label{thm:suff1}
a) Let \eq{kad} and \eq{suff0} hold, and let
\begin{equation}\label{suff11}
\be\ga+\ka_{21}\le 0.
\end{equation}
Then $B_1$ decreases on $[0, +\infty)$.

b) Let \eq{kad}, \eq{suff0} and \eq{suff11} hold, and let
\begin{equation}\label{suffC}
d_0\ge \frac{\al}{2}\ga^2.
\end{equation}
Then $C$ decreases on $[0, +\infty)$, and the bond price is a
decreasing function of $(x, \tau)$ in the region $x_1, x_2>0,
\tau>0$.
\end{thm}
\begin{proof}
a) Set $Y=B'_1$, and solve \eq{B1} w.r.t. $B_1$ taking into
account that the latter is negative:
\[
B_1=\ka_{11}-\sqrt{\ka_{11}^2+2\ka_{12}B_2-\be B_2^2+2d_1+2Y}.
\]
By differentiating \eq{B1}, we find
\begin{eqnarray}\label{Yder}
Y'&=&-\ka_{11}Y+YB_1-\ka_{21}B'_2+\be B_2 B'_2\\\nonumber &=&
-Y\sqrt{\ka_{11}^2+2\ka_{12}B_2-\be B_2^2+2d_1+2Y}+B'_2(\be
B_2-\ka_{21}).
\end{eqnarray}
On the strengh of \eq{kad} and \eq{solB2},  $Y$ is negative in a
small right neighborhood of 0, and for all $\tau>0$,
$B'_2(\tau)<0$ and $B_2(\tau)\in (-\ga, 0)$. Hence, under
condition \eq{suff11}, $B'_2(\tau)(\be B_2(\tau)-\ka_{21})<0,\
\forall \ \tau>0$, and if $Y(\tau)$ tries to approach $0$ from
below, $Y'(\tau)$ becomes negative. Thus, $B'_1=Y$ must remain
negative on $(0, +\infty)  $.

b) The RHS of \eq{C} is less than $-d_0+\al \ga^2/2$.
\end{proof}
The next theorem justifies the use of the formal solution of
 ATSM.
 \begin{thm}\label{thm:reduction1} Let $g$ be bounded and continuous, and let
\begin{equation}\label{suff1}
d_1+\frac{\ka_{11}^2}{2}-\ka_{21}\ga -\frac{\be}{2}\ga^2> 0.
\end{equation}
Then the  expressions \eq{defcc}, \eq{st1} and  \eq{affr} are  the
unique  solution to the problem \eq{fc1}-\eq{bc1} in the class of
continuous functions, which admit the bound
\begin{equation}\label{fbound1}
|f(x, \tau)|\le C\exp[(-\ga x_2)_+].
\end{equation}
\end{thm}
\begin{proof} The general scheme of the proof is described in
\sect{genjust}, and the proof itself is given in
\sect{reductiona1n}.
\end{proof}

\noindent {\em Remark 3.1.} The bound \eq{fbound1}, which
specifies the class of functions among which the Feynman-Kac
formula gives the unique solution, is natural because the bond
price given by \eq{deff0} satisfies this bound: see \eq{solB2}.

\vskip0.2cm \noindent{\em Remark 3.2.}  The LHS in the condition
\eq{suff1} summarizes the influence of different parameters of the
process and the affine model of the short rate, which act in
opposite directions. The main obstacle for the proof of the
Feynman-Kac theorem (and the validity of this theorem) comes from
the region in the state space where the short rate is unbounded
from below. The larger the $d_1$, the smaller this region is, and
the condition \eq{suff1} provides the lower bound for $d_1$, which
ensures the validity of the Feynman-Kac (the other parameters
remaining fixed). In the $A_0(n)$ case, when the volatility is
bounded and so trajectories of the process are not pushed too far
too fast in the ``negative" region, no additional condition is
needed. In the $A_1(n)$ case, the entries of the volatility matrix
are large in the region where $x_1$ is large, therefore the less
time a trajectory of the process spends there, the smaller
obstacle for the proof of the Feynman-Kac theorem (and the
validity of this theorem) is. A large value of the mean-reverting
entry $\ka_{11}$ ensures that trajectories of the process do not
remain far from the line $x_1=0$ for too long time, and the
condition \eq{suff1} provides the lower bound for $\ka_{11}$, the
other parameters remaining fixed.

The parameters $d_2$, $\be$ and $\ka_{22}$ act in opposite
directions as well. The larger the $d_2$, the larger (the absolute
value of) negative values of the short rate can be, and the larger
the $\be$, the farther a trajectory can be pushed into the
``negative" region. The larger the $\ka_{22}$ is, the faster a
trajectory of the process will return to a neighborhood of a line
$x_2=0$, where the short rate is bounded from below. If
$\ka_{21}=0$, then \eq{suff1} provides an upper bound on $\be
(d_2/\ka_{22})^2$, which ensures the validity of the Feynman-Kac
theorem. If $\ka_{21}\neq 0$, the dynamics becomes more complex,
and the  term $\ka_{21}\ga$ takes this additional complexity into
account.

\begin{cor} For the bond pricing problem,  \theor{thm:reduction1} is valid under condition
\begin{equation}\label{suff2}
d_1+\frac{\ka_{11}^2}{2}-\ka_{21}\ga- \frac{\be}{2}\ga^2\ge 0.
\end{equation}
\end{cor}
\begin{proof}
Indeed, suppose that  \eq{suff2} holds with the equality. Then we
can approximate $d_1$ by a sequence $\{d_{1,m}\}$ converging to
$d_1$ from above. Each $d_{1, m}$ satisfies \eq{suff1}, and hence
the conclusion of \theor{thm:reduction1} holds. Set
$r_m(x)=d_0+d_{1,m}x_1+d_2 x_2$. By the Dominant Convergence
Theorem, $f(r_m, g; x, t)\to f(r, g; x, t)$, point-wise, and the
convergence $f_0(r_m, g; x, t)\to f_0(r, g; x, t)$ follows from
the explicit formula for these functions and the theorem about the
continuous dependence
 on parameters of the solution of a system of ODE of the first
order.
\end{proof}

\subsection{Family $A_1(n)$}\label{maina1n}
 The state space is $\R_+\times
\R^{n-1}$, the $r$ is given by \eq{affr} with $d_j\ge 0,
j=1,\ldots, n$, and $d_n>0$, and the infinitesimal generator of
the process is of the form
\begin{equation}\label{gena1n}
L=\langle\theta-\ka x, \dd\rangle+\frac{1}{2}
x_1\dd_1^2+\frac{1}{2}\sum_{j,k=2}^{n}(\al_{jk}+x_1\be_{jk})\dd_j\dd_k,
\end{equation}
where $\theta_1, \ka_{jj}, j=1,\ldots, n$  are positive,
$\ka_{jl}\le 0, 1\le l<j$, $\ka_{jk}=0, j<k$,
 and
$\al=[\al_{jk}]_{j,k=2}^n, \be=[\be_{jk}]_{j,k=2}^n$ are positive
definite matrices (these restrictions can be relaxed). Without
loss of generality, we may assume $\theta_j=0, j\ge 2 $, when
convenient.

Introduce vectors $d^2=(d_2, \ldots, d_n)$,
$\ka^{21}=(\ka_{21},\ldots, \ka_{n1})$,
 and set $\ga=(\ka^T)^{-1}d$.
 The subsystem
of the Riccati equations for $\cB=(B_2,\ldots, B_n)$ enjoys the
same properties as the subsystem for $B$ in the $A_0(n)$-model
above, therefore the same proof as of \theor{necsuf0n} gives
 \begin{lem}\label{lem:bj}
For $j=2,\ldots, n$, function $B_j$ is continuous and
non-increasing on $[0, +\infty)$, and $B_j(0)=0,
B_j(+\infty)=-\ga_j$.
\end{lem}
It remains to study the monotonicity of $B_1$ and $C$. Introduce
 quadratic polynomials
\[
Q_B(y)=-d_1-\langle y, \ka^{21}\rangle +\frac{1}{2}y^T\be y,\quad
y\in\R^{n-1},
\]
and
\[
Q_C(y)=-d_0+\frac{1}{2}y^T\al y,  \quad y\in\R^{n-1}.
\]
The remaining two Riccati equations, for $B_1$ and $C$, are
\begin{eqnarray}\label{B1n}
B'(\tau)&=&-\ka_{11}B_1(\tau)+\frac{1}{2}B_1^2(\tau)+Q_B(\cB(\tau)),\\\label{C1n}
C'(\tau)&=&\theta_1B_1(\tau)+Q_C(\cB(\tau)).
\end{eqnarray}
Each of equations \eq{B1n} and \eq{C1n} is similar to  equation
\eq{solC0n} for $C$ in the $A_0(n)$-model, therefore the same
proof as of \theor{necsuf0n} gives
 \begin{thm}\label{thm:necessn}
 a) If $B_1$ is non-increasing on $[0, +\infty)$, then
 \begin{equation}\label{kadn}
 d_1>0\quad {\rm or}\ \langle d^2, \ka^{21}\rangle<0,
 \end{equation}
 and
\begin{equation}\label{nec0n}
Q_B(-\ga_2,\ldots, -\ga_n)\le 0.
\end{equation}
b) Let \eq{kadn}  hold, and
\begin{equation}\label{suffn0}
Q_B(y)\le 0, \quad \forall\ y\in \{-\ga_2, 0\}\times\cdots\times
\{-\ga_n, 0\}.
\end{equation}
 Then $B_1$ decreases on $[0, +\infty)$.

 c) Let \eq{kadn} and \eq{suffn0} hold, and
  \begin{equation}\label{suffCn}
  Q_C(y)\le 0, \quad \forall\ y\in \{-\ga_2, 0\}\times\cdots\times
\{-\ga_n, 0\}.
\end{equation}
Then $C$ decreases on $[0, +\infty)$, and the bond price is a
decreasing function of $(x, \tau)$ in the region $x>0, \tau>0$.
\end{thm}
 By using the family of transformations described in Dai and
 Singleton (2000), it is possible to reduce any $A_1(n)$-model to a model
 with $\ka$ satisfying
 \begin{equation}\label{ka}
 \ka_{jk}=0,\quad \forall\ k\neq j,\ k\ge 2,\ j\ge 1,
 \end{equation}
  and we will formulate the theorem about the
 justification of the use of the Feynman-Kac theorem under this
 assumption.
 \begin{thm}\label{thm:reductionn} Let $g$ be bounded and continuous,
 let \eq{ka} hold,
 and let
\begin{equation}\label{suffn1}
\frac{\ka_{11}^2}{2}-Q_B(y)> 0, \quad \forall\ y\in \{-\ga_2,
0\}\times\cdots\times \{-\ga_n, 0\}.
\end{equation}
Then the  expressions \eq{defcc}, \eq{st1} and \eq{affr} are  the
unique solution to the problem \eq{fc1}-\eq{bc1} in the class of
continuous functions, which admit the bound
\begin{equation}\label{fboundn}
|f(x, \tau)|\le C\exp\left[\sum_{j=2}^{n}(-\ga_j x_j)_+\right]
\end{equation}
\end{thm}
\begin{proof} In \subsect{partn}. \end{proof}
\noindent {\em Remark 3.3.} The bound \eq{fboundn}, which
specifies the class of functions among which the Feynman-Kac
formula gives the unique solution, is natural because the bond
price given by \eq{deff0} satisfies this bound: see \lemm{lem:bj}.

\vskip0.2cm \noindent{\em Remark 3.4.}   In order to apply
\theor{thm:reductionn} to a particular model, the family of
transformations described in Dai and
 Singleton (2000) should be used to obtain a model which satisfies \eq{ka}.
 The condition \eq{suffn1} is a natural generalization of the condition \eq{suff1}
 for the multi-factor case; the interpretation is essentially the same as in
 Remark 3.2.

\begin{cor} For the bond pricing problem, condition \eq{suffn1} in \theor{thm:reductionn}
can be replaced by a weaker one:
\begin{equation}\label{suffn2}
\frac{\ka_{11}^2}{2}-Q_B(y)\ge 0, \quad \forall\ y\in \{-\ga_2,
0\}\times\cdots\times \{-\ga_n, 0\}.
\end{equation}
\end{cor}
\subsection{Family $A_2(3)$}\label{maina23}
 By using the family of transformations described in Dai and
 Singleton (2000), it is possible to reduce any three-factor ATSM with two
 factors of the
 CIR-type to a model of the form
 \begin{equation}\label{st23}
 dX(t)=(\theta - \kappa X(t))dt +\sqrt{[S_{jj}(t)]}d B(t),
 \end{equation}
 where $\theta\in \R^3$ is the vector with components $\theta_3=0$,
 \begin{equation}\label{th}
 \theta_1>0,\quad \theta_2>0,
 \end{equation}
 and the entries of  matrix $\ka=[\ka_{jl}]$ satisfy
 \begin{eqnarray}\label{k1}
 \ka_{11}, \ka_{22}, \ka_{33}&>&0,\\\label{k2}
 \ka_{21}, \ka_{12}&\le & 0,\\\label{k3}
 \ka_{11}\ka_{22}-\ka_{12}\ka_{21}&>&0,\\\label{k4}
 \ka_{13}=\ka_{23}&=&0.
 \end{eqnarray}
Further,
 \begin{eqnarray}\label{S1}
 S_{jj}(t)&=&\be_{jj}X_j(t), \quad j=1,2,\\\label{S3}
 S_{33}(t)&=&\al_3 + \be_{31}X_1(t) + \be_{32} X_2(t),
 \end{eqnarray}
 where
 \begin{equation}\label{abe}
 \al_3, \be_{11}, \be_{22}>0,\quad \be_{31}, \be_{32}\ge 0,
 \end{equation}
 and finally, the short rate process is given by \eq{affr} with
 \begin{equation}\label{d}
 d_0\in\R,\quad d_1, d_2\ge 0, \quad d_3>0.
 \end{equation}
 Notice that under conditions \eq{th} and \eq{k1}--\eq{k4}, any
 trajectory of the process $X$, which starts in the region $x_1\ge
 0,\ x_2\ge 0$, remains in this region, a.s.

 The infinitesimal generator of the process is
 \[
 L=(\theta -\ka x)^T\dd_x
 +\frac{1}{2}\sum_{j=1,2}\be_{jj}x_j \dd_j^2
 +\frac{1}{2}(\al_3+\be_{31}x_1+\be_{32}x_2)\dd_3^2,
 \]
 and therefore the system of Riccati equations is
 \begin{eqnarray}\label{ri1}
 B'_1&=&-\ka_{11}B_1-\ka_{21}B_2+\frac{\be_{11}}{2}B_1^2-d_1
 -\ka_{31}B_3+\frac{\be_{31}}{2}B_3^2,\\\label{ri2}
 B'_2&=&-\ka_{12}B_1-\ka_{22}B_2+\frac{\be_{22}}{2}B_2^2-d_2
 -\ka_{32}B_3+\frac{\be_{32}}{2}B_3^2,\\\label{ri3}
 B'_3&=&-\ka_{33}B_1-d_3,\\\label{ri4}
 C'&=&-d_0+\theta_1 B_1 +\theta_2 B_2+
 \frac{\al_3}{2}B_3^2.
 \end{eqnarray}
 From the initial condition
 \begin{equation}\label{bc23}
 B_1(0)=B_2(0)=B_3(0)=C(0)=0
 \end{equation}
 and \eq{ri3} we can easily find $B_3$;
 \begin{equation}\label{solB3}
 B_3(\tau)=-\ga(1-e^{-\ka_{33}\tau}),
 \end{equation}
 where $\ga=d_3/\ka_{33}>0$. Clearly, $B_3$ decays on $[0,
 +\infty)$, and
 \begin{eqnarray}\label{B31}
 B_3(+\infty)&=&-\ga,\\\label{B32}
 B_3(\tau)&\in& (-\ga, 0),\quad \forall\ \tau>0,\\\label{B33}
 B_3(\tau)&\sim &-d_3\tau, \quad {\rm as}\ \tau\to+0.
 \end{eqnarray}
 The monotonicity of $B_j, j=1,2,$ implies additional restrictions
 on the parameters of the model. To formulate them,
 denote by $\tka^{11}$ the inverse to the upper left $2\times 2$ block
 $\ka^{11}:=[\ka_{jl}]_{j,l=1,2}$ of matrix $\ka$,
 \[
 \tka^{11}=\frac{1}{\ka_{11}\ka_{22}-\ka_{12}\ka_{21}}\left[
 \begin{array}{rr}
 \ka_{22} & -\ka_{12}\\
 -\ka_{21} & \ka_{11}
 \end{array}\right]
 \]
 and set
 \[
 \left[\begin{array}{r}
 \td_1\\
 \td_2\end{array}\right]=\tka^{11}\left[\begin{array}{r}
 d_1\\
 d_2\end{array}\right],\quad
 \left[\begin{array}{r}
 \tka_{31}\\
 \tka_{32}\end{array}\right]=\tka^{11}\left[\begin{array}{r}
 \ka_{31}\\
 \ka_{32}\end{array}\right],\quad
 \left[\begin{array}{r}\tbe_{31}\\
 \tbe_{32}\end{array}\right]=\tka^{11}\left[\begin{array}{r}
 \be_{31}\\
 \be_{32}\end{array}\right].
 \]
 On the strength of \eq{k1}--\eq{k3}, the diagonal (resp.,
 off-diagonal) entries of $\tka^{11}$ are positive (resp.,
 non-positive), therefore
 \begin{equation}\label{tp}
 \tbe_{31}\ge 0,\quad \tbe_{32}\ge 0,
 \end{equation}
 and
 \begin{eqnarray}\label{td}
 d_1, d_2\ge 0 & \Rightarrow & \td_1, \td_2\ge 0,\\\label{tka}
\ka_{31}, \ka_{32}< 0 & \Rightarrow & \tka_{31}, \tka_{32}< 0.
\end{eqnarray}
 \begin{thm}\label{thm:necess23}
 Let $B_j, j=1,2,$ be non-increasing on $[0, +\infty)$. Then the
 following conditions hold, for $j\neq l\in \{1, 2\}$:
 \begin{eqnarray}\label{nec211}
 d_j>0 \quad {\rm or}\quad d_j=0 \ &{\rm and}&\
 \ka_{lj}d_l+\ka_{3j}d_3\le 0;\\\label{nec212}
\td_j>0 \quad {\rm or}\quad \td_j=0 \ &{\rm and}&\
 \tka_{3j}\le 0;\\\label{nec213}
 \td_j-\tka_{3j}\ga-\frac{\tbe_{3j}}{2}\ga^2&\ge & 0.
 \end{eqnarray}
 \end{thm}
\begin{proof} a) If $d_j<0$, then \eq{ri1}--\eq{ri2} and \eq{bc23} imply
that $B'_j(\tau)>0$ in a right neighborhood of 0, contradiction.
Hence, both $d_j\ge 0$. If both $d_j=0$, then from
\eq{ri1}--\eq{ri2} and \eq{B33}, both $B_j(\tau)=o(\tau)$, as
$\tau\to +0$, and moreover,
\[
B'_j(\tau)\sim \ka_{3j}d_3\tau,\quad \tau\to +0.
\]
Hence, it is necessary that $\ka_{3j}\le 0$. Finally, if $d_1=0$
but $d_2>0$, then $B_2(\tau)\sim - d_2\tau$, as $\tau\to +0$,
hence,
\[
B'_1(\tau)\sim (\ka_{21}d_2+\ka_{31}d_3)\tau, \quad \tau\to +0.
\]
This excludes the case $\ka_{21}d_2+\ka_{31}d_3>0$, and finishes
the proof of $\eq{nec211}$.

b) To prove \eq{nec212}--\eq{nec213}, we apply $\tka^{11}$ to
subsystem \eq{ri1}--\eq{ri2}; the result is
\begin{eqnarray}\label{ri11}
\tka^{11}\left[\begin{array}{r} B'_1\\ B'_2\end{array} \right] &=&
- \left[\begin{array}{r} B_1\\ B_2\end{array} \right]
+\frac{1}{2}\tilde B\left[\begin{array}{r} B_1^2\\
B_2^2\end{array} \right]\\\nonumber
&&-\left[\begin{array}{r} \td_1\\
\td_2\end{array} \right]- B_3\left[\begin{array}{r}
\tka_{31}\\
\tka_{32}\end{array} \right]+\frac{1}{2}B_3^2\left[\begin{array}{r} \tbe_{31}\\
\tbe_{32}\end{array} \right],
\end{eqnarray}
where
\[
\tilde B = \tka^{11}\left[\begin{array}{ll} \be_{11} & 0\\
0 & \be_{22}\end{array}\right].
\]
Since $B'_1$ and  $B'_2$ are non-positive, and the entries of
$\tka^{11}$ are non-negative, the LHS in \eq{ri11} is
non-positive. Now by using \eq{ri11} and arguing as in part a)
above, we deduce \eq{nec212}.

If \eq{nec213} fails, then from \eq{B31} we conclude that in a
neighborhood of $+\infty$, one of the components of the RHS in
\eq{ri11} is positive, contradiction.
\end{proof}
 The next theorem gives sufficient conditions for the monotonicity
 of $B_j$ and $C$; they are more stringent than the necessary
 conditions in \theor{thm:necess23}.
\begin{thm}\label{thm:suff23}
a) Let the following conditions hold, for $j=1,2$:
\begin{eqnarray}\label{suff211}
d_j>0 \quad {\rm or}\quad d_j=0\ {\rm and}\
\ka_{3j}<0;\\\label{suff212}
d_j-\ka_{3j}\ga-\frac{\be_{3j}}{2}\ga^2\ge 0;\\\label{suff213}
\ka_{3j}+\be_{3j}\ga\le 0,\\\label{beka} \be_{3j}>0 \quad {\rm or}
\quad \ka_{3j}<0.
\end{eqnarray}
Then $B_1$ and $B_2$ are decreasing on $[0, +\infty)$.

b) In addition, let \begin{equation}\label{suff2C} d_0\ge
\frac{\al_3}{2}\ga^2.
\end{equation}
Then $C$ decreases on $[0, +\infty)$, and the bond price is a
decreasing function of $(x, \tau)$ in the region $x_1>0, x_2>0,
\tau>0$.
\end{thm}
\begin{proof}
a) First, we show that there exists $\tau_0>0$ such that for
$j=1,2,$
\begin{equation}\label{negBj}
B_j(\tau)<0,\quad 0<\tau< \tau_0,
\end{equation}
and
\begin{equation}\label{negBpj}
B'_j(\tau)<0,\quad  0<\tau< \tau_0.
\end{equation}
We use \eq{ri1}--\eq{ri2}, \eq{bc23} and \eq{B33}. If $d_1$ and
$d_2$ are positive, then $B'_j(\tau)\sim -d_j$, as $\tau\to +0$,
hence \eq{negBpj} hold, and \eq{negBj} holds as well. If both
$d_j=0$ but both $\ka_{3j}<0$, then $B'_j(\tau)\sim
\ka_{3j}d_3\tau$, as $\tau\to +0$, and \eq{negBj} and \eq{negBpj}
hold. Finally, if one of $d_j$, say, $d_2$, is positive, and the
other, $d_1$, is 0, then $B'_2(\tau)\sim -d_2$, $B_2(\tau)\sim
-d_2\tau$, as $\tau\to +0$, and
\[
B'_1(\tau)\sim (\ka_{21}d_2+\ka_{31}d_3)\tau,\quad \tau\to +0.
\]
Since $\ka_{21}\le 0, d_2\ge 0$ and $d_3>0$, we obtain
\eq{negBj}--\eq{negBpj}.

Second, we show that \eq{negBj} holds with $\tau_0=+\infty$. Under
condition \eq{suff211}, the polynomials
\[
Q_j(y)=-d_j-\ka_{3j}y+\frac{\be_{3j}}{2}y^2
\]
are negative on $(-\ga, 0)$, therefore in view of \eq{B32},
\begin{equation}\label{dj}
-d_j+\ka_{3j}B_3(\tau)+\frac{\be_{3j}}{2}B_3(\tau)^2<0,\quad
\forall\ \tau>0.
\end{equation}
Suppose that as $\tau$ increases, $B_j(\tau), j=1,2,$ start to
approach 0 from below, simultaneously. Then from
\eq{ri1}--\eq{ri2} and \eq{dj}, at least one of $B'_j(\tau)$
becomes negative before both $B_j(\tau), j=1,2,$ reach 0,
contradiction. If $B_1(\tau)$ is approaching 0 but $B_2(\tau)$ is
not, then eventually, on the strength of \eq{ri1} and condition
$\ka_{21}\le 0$, $B'_1(\tau)$ becomes negative, contradiction.
Thus, \eq{negBj} holds on the whole half-axis.

It remains to prove that $Y_j:=B'_j$ are negative on the whole
half-axis, $j=1,2$. We differentiate \eq{ri1}--\eq{ri2}:
\begin{eqnarray}\label{Y1}
Y'_1&=&-\ka_{11}Y_1-\ka_{21}Y_2 +\be_{11}B_1Y_1-\ka_{31}Y_3
+\be_{31}B_3Y_3,\\\label{Y2}
 Y'_2&=&-\ka_{12}Y_1-\ka_{22}Y_2
+\be_{22}B_2Y_2-\ka_{32}Y_3 +\be_{32}B_3Y_3,
\end{eqnarray}
and rewrite \eq{Y1}--\eq{Y2} as
\begin{eqnarray*}
Y'_1&=&-(\ka_{11}-\be_{11}B_1)Y_1-\ka_{21}Y_2
+(\be_{31}B_3-\ka_{31})Y_3,\\
 Y'_2&=&-\ka_{12}Y_1-(\ka_{22}-\be_{22}B_2)Y_2
+(\be_{32}B_3-\ka_{32})Y_3.
\end{eqnarray*}
Since $B_j$ are negative, $\ka_{jj}-\be_{jj}B_j>0, j=1,2$, and due
to \eq{B32}, \eq{beka} and \eq{suff213}, all the expressions in
the brackets are positive.  Since $Y_3$ is negative, $Y'_1(\tau)$
and $Y'_2(\tau)$ cannot approach 0 from below simultaneously:
indeed, then eventually, the RHS's will become negative,
contradiction. Since $-\ka_{21}\ge 0$, we have
$-\ka_{21}Y_2(\tau)\le 0$ where $Y_2(\tau)<0$, therefore by the
same reasoning, $Y_1(\tau)$ cannot approach 0 from below while
$Y_2(\tau)$ remains separated from 0. By interchanging the indices
1 and 2, we conclude that \eq{negBpj} holds on the whole
half-axis.

b) Recall that $\theta_1$ and $\theta_2$ are positive, and apply
\eq{ri4}, \eq{B32} and \eq{negBj}.
\end{proof}

\begin{thm}\label{thm:reduction23} Let $g$ be bounded and continuous, and let
the following conditions hold: for $y=0, \ga$,
\begin{eqnarray}\label{suff21}
d_1+\frac{\ka_{11}^2}{2\be_{11}}+\frac{\ka_{21}\ka_{22}}{\be_{22}}-\ka_{31}y
-\frac{\be_{31}}{2}y^2&>&0,\\\label{suff22}
d_2+\frac{\ka_{22}^2}{2\be_{22}}+\frac{\ka_{12}\ka_{11}}{\be_{11}}-\ka_{32}y
-\frac{\be_{32}}{2}y^2&>&0.
\end{eqnarray}
Then the  expressions \eq{defcc}, \eq{st1} and \eq{affr} are  the
unique  solution to the problem \eq{fc1}-\eq{bc1} in the class of
continuous functions, which admit the bound
\begin{equation}\label{fbound21}
|f(x, \tau)|\le C\exp[(-\ga x_3)_+].
\end{equation}
\end{thm}
\begin{proof} In \subsect{reduction23}. \end{proof}
\noindent {\em Remark 3.5.} The bound \eq{fbound21}, which
specifies the class of functions among which the Feynman-Kac
formula gives the unique solution, is natural because the bond
price given by \eq{deff0} satisfies this bound: see \eq{solB3}.

\vskip0.2cm \noindent{\em Remark 3.6.}
 The pair of conditions \eq{suff21}-\eq{suff22} is a natural generalization of the condition \eq{suff1}
 in the $A_1(2)$ model; the interpretation is essentially the same as in
 Remark 3.2.

\section{Justification of the use of the Feynman-Kac formula}
\label{genjust}

\subsection{General scheme}\label{genscheme}
 {\em Step 1.} We
assume that $g$ is non-negative, bounded and sufficiently regular,
so that in the case of a continuous $r$ bounded from below (not
necessarily affine), problem \eq{fc1}-\eq{bc1} has a  solution,
$f_0(g, r; \cdot, \cdot)$, in the class of bounded continuous
functions, which is given by the stochastic expression \eq{defcc}:
$f_0(g, r; x, t)=f(g, r; x, t)$ for all $x$ and $t<T$. For affine
diffusions, the coefficients of the infinitesimal generator of the
process satisfy the global Lipschitz condition, and hence, the
Feyman-Kac theorem is applicable. (The reduction to the case of
more general $g$ is fairly standard).

{\em Step 2.} Fix $N\in\R$, and for the affine $r$, set
$r_N(x)=\max\{r(x), -N\}$. Then $r_N$ is bounded from below, hence
both $f(g, r_N; x, t)$ and $f_0(g, r_N; x, t)$ exist, and $f(g,
r_N; x, t)=f_0(g, r_N; x, t)$ for all $x$ and $t<T$.

{\em Step 3.} By the Monotone Convergence Theorem,
\[
f(g, r_N; x, t)\to f(g, r; x, t)\quad {\rm as}\ N\to +\infty,
\]
point-wise, therefore it remains to show that
\begin{equation}\label{conv}
f_0(g, r_N; x, t)\to f_0(g, r; x, t)\quad {\rm as}\ N\to +\infty,
\end{equation}
point-wise.

{\em Step 4.} We prove \eq{conv} by reducing to the case of a
family of short rates, which is bounded from below uniformly in
$N$. To this end, we take a non-negative function $\phi\in
C^\infty(\rn)$, and consider the representation
\begin{equation}\label{conj}
\exp(-\phi(x))L\exp(\phi(x))=L_\phi-\tr_\phi(x),
\end{equation}
where $L_\phi$ is a differential operator without the zero-order
term, and $\tr_\phi$ is a function. In the presence of factors of
the CIR-type, the infinitesimal generator may have non-trivial
affine coefficients at the second order derivatives, therefore the
choice of $\phi$ in the form of a quadratic polynomial, as in the
proof of \theor{fka0n}, is impossible except for some very special
cases. One may try $\phi$'s which are (approximately) positive
homogenous of degree 1 in a neighborhood of infinity. The simplest
version $\phi(x)=\langle Ax, x\rangle/\langle x\rangle$, where
$\langle x\rangle:=(1+||x||^2)^{1/2}$, is possible but it requires
unnecessary strong restrictions on parameters of the model. It
turns out that the construction of $\phi$ should be adjusted to
each model. For the proof to work, the following general
properties of $\phi$ are essential (they can be relaxed, though):
\begin{enumerate}[(i)]
\item
there exist constants $c_0>0$ and  $C$ such that
\begin{equation}\label{phibound}
\phi(x)\ge c_0|x|-C;
\end{equation}
\item
there exist constants $C_0>0, M, C_1$ such that $r_{\infty,
\phi}:=r+\tr_\phi$  admits  the following estimate:
\begin{equation}\label{rbound}
 C_0|x|-M\le
r_{\infty, \phi}(x)\le C_1|x|+M, \quad \forall\ x;
\end{equation}
\item
there exist constants $C_2$ and $\de>0$ such that for all $x$ and
$t\in [0, T]$, the function $f_{\phi}(g, r; x,
t):=\exp(-\phi(x))f_0(g, r; x, t)$ satisfies an estimate
\begin{equation}\label{f0bound}
|f_{\phi}(g, r; x, t)|\le C_2\exp(-\de |x|);
\end{equation}
\item
$L_\phi$ is sufficiently regular in the sense that
 for any $r^0$ which admits the bound \eq{rbound}, and a continuous $g^0(x)$
  admitting the bound
\begin{equation}\label{g0bound}
|g^0(x)|\le C_3\exp(-\de |x|),
\end{equation}
where $\de>0$ and $C_3$ are independent of $x$,
 a continuous
solution to the problem
\begin{eqnarray}\label{fc2}
(\pdt +L_\phi -r^0)f(x, t)&=&0,\quad 0\le t<T;\\\label{bc2} f(x,
T)&=&g^0(x),
\end{eqnarray}
which exponentially decays at infinity, exists and it is unique;
call it $f^\phi(g^0, r^0; x, t)$ (we add indices 0, and use labels
$r^0$ and $g^0$) in order to avoid the confusion with $g$ and $r$
in \eq{fc1}-\eq{bc1});
\item
as $N\to +\infty$,
\begin{equation}\label{conv2}
f^\phi(g^0, r_{N, \phi}; x, t)\to f^\phi(g^0, r_{\infty, \phi}; x,
t),
\end{equation}
point-wise; here $r_{N, \phi}:=r_N+\tr_\phi$.
\end{enumerate}
Notice that the construction of $\phi$ will depend on a small
parameter $\eps\in (0, 1)$, and the $r_{N, \phi}$ will satisfy a
weak version of the global Lipschitz condition, with parameter,
which simplifies the proof of the existence and uniqueness theorem
in \sect{reductiona1n}. We do not specify this condition here
because it is a useful technical tool only, and it should be
possible to prove the existence and uniqueness theorem under
weaker conditions.

When the existence and uniqueness theorem is proved, the
convergence \eq{conv2} is typically not difficult to establish.
For instance, if it can be shown that $L_\phi$ is the generator of
a Markov process without killing, and the Feynman-Kac theorem is
applicable, then \eq{conv2} can be easily deduced from the
Feynman-Kac formula and the Dominant Convergence Theorem. However,
for the $L_\phi$, which we will construct below, there is no ready
Feynman-Kac theorem available, and so \eq{conv2} will be proved
differently, by using the representation theorem for analytic
semigroups.

When the $\phi$ is constructed, and properties (i)--(v) are
established, we can prove \eq{conv} as follows. For $N=1,2,\ldots,
\infty$, set
\begin{eqnarray*}
g_\phi(x)&=&\exp(-\phi(x))g(x),\\
 f_\phi(g, r_N; x, t)&=&\exp(-\phi(x))f_0(g,
r_N; x, t),
\end{eqnarray*}
substitute $f_0(g, r_N; x, t)=\exp(\phi(x))f_\phi(g, r_N; x, t)$
into \eq{fc1}-\eq{bc1}, and multiply by $\exp(-\phi(x))$; we
obtain that $f_\phi(g, r_N; \cdot, \cdot)$ is a continuous
solution to the problem
\begin{eqnarray}\label{fc3}
(\pdt +L_\phi -r_{N,\phi})f(x, t)&=&0,\quad 0\le t<T;\\\label{bc3}
f(x, T)&=&g_\phi(x),
\end{eqnarray}
which exponentially decays at infinity.
 Clearly, $r_{N,\phi}$ satisfy \eq{rbound}
with the constants independent of $N=1,2,\ldots, \infty$;  the RHS
in \eq{bc2} admits bound \eq{g0bound} since $g$ is bounded, and
$\phi$ satisfies \eq{phibound}. By (iv), the continuous solution
to problem \eq{fc3}--\eq{bc3}, which exponentially decays at
infinity, is unique, hence $f_\phi(g, r_N; \cdot,
\cdot)=f^\phi(g_\phi, r_{N,\phi}; \cdot, \cdot)$, and by (v), as
$N\to+\infty$, $f^\phi(g_\phi, r_{N,\phi}; \cdot, \cdot)$
converges to $f^{\phi}(g_\phi, r_{\infty, \phi}; \cdot, \cdot)$,
pointwise; \eq{conv} follows.

\subsection{Outline of the verification of conditions (iv)--(v)}\label{outline}
Let $x'$ be the group of the CIR-type variables, and
$x^{\prime\prime}$ the group of the other variables; then $x=(x',
x^{\prime\prime})$. Let $(R_+)^m\times\R^{n-m}$ be the
corresponding decomposition of the state space. We apply the
standard approach: first, we write problem \eq{fc3}--\eq{bc3} in
the form
\begin{eqnarray}\label{fcop}
F'(\tau)+A F(\tau)&=&0,\quad \tau>0;\\\label{bcop}
 F(+0)&=& g^0,
 \end{eqnarray}
 where $A= -L_\phi+r^0$ is a (partial) differential operator on
$(R_+)^m\times\R^{n-m}$, and for each $\tau$, $F(\tau)$ is a
function on $(R_+)^m\times\R^{n-m}$. Next,
 by using the Laplace
transform w.r.t. $\tau$, we reduce problem \eq{fc3}--\eq{bc3} to
the family of problems
 \begin{equation}\label{fcl}
 (\la+A)\hF(\la)=g^0,
 \end{equation}
where $\hF$ is the Laplace transform of $F$, and $\la$ belongs to
a half-plane of the complex plane, of the form $\{\la\ |\ \Im
\la>\la_0\}$. Then, by using the theory of degenerate elliptic
operators with parameter, we show that if $\la_0$ is sufficiently
large,  then the operator $\la+A$ is invertible uniformly w.r.t.
$\la$ in the half-plane $\Im\la\ge \la_0$. This proves the
existence and uniqueness of the solution to the boundary problem
\eq{fc3}-\eq{bc3} (that is, part (iv) of the general scheme) but
in a wider class of functions since under this approach, \eq{bcop}
is satisfied in a weak sense (in the sense of the theory of
generalized functions). To show that the solution satisfies
\eq{bcop} in the strong sense, and to prove the convergence in
part (v), some additional effort is needed. We  show that the
representation theorem for analytic semigroups can be applied to
$A$, and derive \eq{conv2} from this representation. We need the
following definition and theorem.

For $\sg\in (0, \pi)$ and $\la_0\ge 0$, set $\Sigma_{\sg,
\la_0}=\{\la\in\C\ |\ |\la|\ge \la_0,\ \arg \la\in [-\sg, \sg]\}$.
Let $A$ be an operator in the Banach space $\cB$, and let there
exist $\sg\in (0, \pi)$, $\la_0\ge 0$, and $C_1$ such that for
$\la\in  \Sigma_{\sg, \la_0}$, $\la+A$ is invertible, and the
resolvent satisfies the estimate
\begin{equation}\label{resolvbound}
||(\la+A)^{-1}||\le C_1(1+|\la|)^{-1}.
\end{equation}
Then $A$ is called a {\it weakly $\sg$-positive
operator}\footnote{Usually, the label $\theta$ is used instead of
$\sg$; unfortunately, $\theta$ is already occupied as the standard
notation in ATSM's}. If $C=0$, $A$ is called {\it $\sg$-positive}.
Let $\LL_{\sg, \la_0}=\partial \Sigma_{\sg, \la_0}$ be a regular
contour, with a parameterization $\la=\la(t)$ satisfying
$\arg\la(t)=\pm\sg$ for $t$ in a neighborhood of $\pm\infty$.

The following theorem is a special case of the representation
theorem for the analytic semigroups (see Section IX.10 in Yosida
(1964)).
\begin{thm}\label{thyosida}
Let $A$ be an unbounded operator in the Banach space $\cB$, and
let there exist $\sg\in (\pi/2, \pi)$ and $C$ such that $A$ is
weakly $\sg$-positive.

  \noindent
  Then
\begin{enumerate}[a)]
\item
 for any $g^0\in \cB$, and any $\tau>0$,
  the following integral is well-defined
 \begin{equation}\label{solyosida}
\exp(-\tau A)g^0=(2\pi i)^{-1}\int_{\LL_{\sg, \la_0}} e^{\tau\la}
(\la+A)^{-1}g^0d\la;
\end{equation}
\item
 \eq{solyosida} defines a strongly continuous semigroup $\{T_\tau\}_{\tau\ge 0}$ in
 $\cB$ by
   $T_0=I$, $T_\tau=\exp(-\tau A)$, $\tau>0$,  and
  \item
for any $g^0\in \cB$, $F(\tau)=\exp(-\tau A)g^0$ is a strongly
continuous solution to the problem \eq{fcop}--\eq{bcop}.
\end{enumerate}
\end{thm}
 As $\cB$, we take $L_{2}((R_+)^m\times\R^{n-m})$,
and we show that
 the conditions of \theor{thyosida} are satisfied. For a function $h$, denote by $h(\cdot)$
 or simply $h$ the-multiplication-by-$h$-operator.
We  will be able to prove that if $\rho>0$ is sufficiently small
then
\begin{equation}\label{estinv}
||\exp(\rho
\langle\cdot\rangle^{\prime\prime})(\la+r^0(\cdot))(\la+A)^{-1}\exp(-\rho
\langle\cdot\rangle^{\prime\prime})||\le C_1,
\end{equation}
for all $\la\in \Sigma_{\sg, \la_0}$. By using the representation
\eq{solyosida} and estimate \eq{estinv}, we can prove \eq{conv2}
as follows. Set $r^0=r_{\infty, \phi}, A=-L_\phi+r_{\infty,
\phi}$, and let $F$ be the solution to \eq{fcop}--\eq{bcop}. Let
$F_{(N)}$ be the solution of the problem \eq{fcop}--\eq{bcop} with
$A_{(N)}=-L_\phi+r_{N,\phi}$ instead of $A$. We show that $A$ and
each $A_{(N)}$ satisfies \eq{resolvbound} uniformly in $N$, with
the same $\Sigma_{\sg, \la_0}$, and therefore the representation
\eq{solyosida} holds for $F_{(N)}$ with $A_{(N)}$ instead of $A$.
Notice that
\begin{eqnarray*}
(\la+A_{(N)})^{-1}g^0-(\la+A)^{-1}g^0&=&(\la+A_{(N)})^{-1}(r_{\infty,\phi}-r_{N, \phi})(\la+A)^{-1}g^0\\
&=&(\la+A_{(N)})^{-1}b_{N, \la} C_\la g^1,
\end{eqnarray*}
where
\[
b_{N, \la}:=(r_{\infty,\phi}-r_{N,
\phi})\exp(-\rho\langle\cdot\rangle ^{\prime\prime}
)(\la+r_{\infty, \phi})^{-1},
\]
\[
C_\la=\exp(\rho
\langle\cdot\rangle^{\prime\prime})(\la+r_{\infty,\phi}(\cdot))(\la+A)^{-1}\exp(-\rho
\langle\cdot\rangle^{\prime\prime} ),
\]
and
$g^1:=\exp(\rho \langle\cdot\rangle^{\prime\prime} )g^0$
is a continuous function which decays at infinity, if $\rho< \de$.
 It is easily seen that
 \[
 \lim_{N\to+\infty}\sup_{\la, x}|b_{N, \la}(x)|=0,
 \]
 therefore the norm of the-multiplication-by-$b_{N, \la}$ operator tends to zero as
 $N\to+\infty$, uniformly in $\la\in \LL_{\sg, \la_0}$; the
 norm of $C_\la$ is uniformly bounded w.r.t. $\la\in \LL_{\sg, \la_0}$
 on the strength of
 \eq{estinv}. Hence,
\begin{eqnarray*}
F_{(N)}(\tau)-F(\tau)&=& \exp(-\tau A_{(N)})g^0-\exp(-\tau
A)g^0\\
&=&(2\pi i)^{-1}\int_{\LL_{\sg, \la_0}} e^{\tau\la}
\left((\la+A_{(N)})^{-1}-(\la+A)^{-1}\right)g^0d\la\\
&=&(2\pi i)^{-1}\int_{\LL_{\sg, \la_0}} e^{\tau\la}
(\la+A_{(N)})^{-1}b_{N, \la} C_\la g^1d\la
\end{eqnarray*}
vanishes as $N\to +\infty$, and \eq{conv2} is proved.

The regularity theorem for locally elliptic operators guarantees
that the solutions $f_{(N)}(x, \tau)=F_{(N)}(x)$ are not only in
$L_2((\R_+)^m\times \R^{n-m})$ but continuous as well, and they
decay at the infinity. Hence, the convergence in $C(\R_+;
L_2((\R_+)^m\times \R^{n-m})$, the space of continuous
vector-functions with values in $L_2((\R_+)^m\times \R^{n-m})$,
implies the pointwise convergence of $f_{(N)}(x, \tau)$ to
$f_{(\infty)}(x, \tau)$.

In \sect{reductiona1n}, the construction of $\phi$ and the proof
of the existence and  uniqueness theorem and bound \eq{estinv}
will be provided for the family $A_1(2)$, and then for families
$A_1(n)$, $n\ge 2$, when the construction of $\phi$ becomes more
involved. In \subsect{reduction23}, the modifications of the proof
for the family $A_2(3)$ are outlined, and the proof for other
families $A_n(m)$ is essentially the same.

\section{Proofs of sufficient conditions for the Feynman-Kac
theorem}\label{reductiona1n}
\subsection{Proof of \theor{thm:reduction1}, part I: construction of $\phi$ and
verification of conditions \eq{conj}--\eq{f0bound}}\label{part1}
 Fix any positive
 $\gap$ and
 \begin{equation}\label{gam}
 \gam<-\ga.
 \end{equation}
Next, take a non-decreasing function $\chi\in C^\infty(\R)$ such
that
\begin{equation}\label{chim}
\chi(y)=\gam,\quad y<-1;
\end{equation}
\begin{equation}\label{chip}
\chi(y)=\gap,\quad y>0;
\end{equation}
and
\begin{equation}\label{chibound}
\gam\le \chi(y)\le \gap,\quad \forall\ y\in\R.
\end{equation}
Then, for any $\eps\in (0, 1)$, construct functions
$\chi_\eps(y)=\chi(\eps y)$ and
\[
\psi_\eps(y)=\int_0^y \chi_\eps(s)ds,
\]
and finally, set
\[
\phi_\eps(x)=\ka_{11} x_1 + \psi_\eps(x_2).
\]
Clearly, $\phi_\eps$ satisfies \eq{phibound}. Set
$\de:=\max\{\gap, \ka_{11}, -\ga-\gam\}$; it is positive since
$\ka_{11}>0, \gap>0$ and \eq{gam} holds.
\begin{lem}\label{l:boundf0}
For any $\eps\in (0, 1)$, there exists $C_\eps$ such that for all
$x_1>0$ and  $x_2\in \R$,
\begin{equation}\label{f0bound1} \exp(-\phi_{\eps}(x_2))|f_0(r,
g; x, t)|\le C_\eps\exp(-\de(x_1+|x_2|)).
\end{equation}
In particular, \eq{f0bound} holds. \end{lem} \begin{proof} From
the explicit solution to the bond pricing problem, we know that
$f_0(r, 1; x, t)$ admits the bound \eq{fbound1}. Since $g$ is
bounded, $f_0(r, g; x, t)$ admits the same bounds (with a
different $C$, perhaps). Hence, on the set $\{x\ |\ x_1>0,
x_2>-1/\eps\}$, function $f_0(r, g; \cdot, \cdot)$ is bounded, and
since $\phi_\eps$ is bounded from below by a linear function with
positive coefficients, $\ka_{11}x_1+\gap x_2$, we conclude that
estimate \eq{f0bound1} holds on this set. On the set $\{x\ |\
x_1>0, x_2<-1/\eps\}$, function $f_0(r, g; \cdot, \cdot)$ is
bounded from above by an exponential function of the form
$C\exp(-\ga x_2)$, and $\phi_\eps$ is bounded from below by an
affine function $\ka_{11}x_1+\gam x_2 + c$, where
\[ c=\eps^{-1}\int_0^{-1}\chi(y)dy.\]
In view of our choice \eq{gam}, estimate \eq{f0bound1} holds on
this set as well.
\end{proof}
Now we can check that for any sufficiently small positive $\gap$
and $\eps\in (0, 1)$, and $\gam$ in a sufficiently small left
vicinity of $-\ga$, the function $\phi_\eps$ satisfies conditions
\eq{conj} and \eq{rbound}. The dependence on $\eps$ will also be
used to check conditions (iv)-(v) of the general scheme: it is
convenient to use the dependence on a small parameter, and prove
the existence and uniqueness of the solution of the boundary
problem \eq{fc2}-\eq{bc2} for sufficiently small $\eps>0$.

 We have
\begin{eqnarray*}
\exp(-\phi_\eps(x))L\exp(\phi_\eps(x))&=&(\theta_1-\ka_{11}x_1)(\dd_1+\ka_{11})\\
&&-(\ka_{21}x_1+\ka_{22}x_2)
(\dd_2+\chi_\eps(x_2))\\&&+\frac{1}{2}
x_1(\dd_1+\ka_{11})^2+\frac{\al+\be
x_1}{2}(\dd_2+\chi_\eps(x_2))^2,
\end{eqnarray*}
therefore
\begin{eqnarray}\label{lphi}
L_\phi&=&\theta_1\dd_1+\frac{1}{2} x_1\dd_1^2\\\nonumber
&&+[-\ka_{21}x_1-\ka_{22}x_2 +(\al+\be x_1)\chi_\eps(x_2)]\dd_2
+\frac{\al+\be x_1}{2}\dd_2^2,
\end{eqnarray}
and
\begin{eqnarray}\label{rphi}
r_{\infty, \phi}(x)&=&d_0-\theta_1\ka_{11}-\frac{\al}{2}
(\eps\chi'_\eps(x_2)+\chi_\eps(x_2)^2)\\ \nonumber
&&+[d_1+\frac{\ka_{11}^2}{2}+\ka_{21}\chi_\eps(x_2)-\frac{\be}{2}(\eps\chi'_\eps(x_2)+\chi_\eps(x_2)^2)]x_1
\\
\nonumber &&+[d_2+\ka_{22}\chi_\eps(x_2)]x_2.
\end{eqnarray}
Here $\chi'_\eps(y):=(\chi')_\eps(y):=\chi'(\eps y)$.
\begin{lem}\label{l:bounrdphi}
Let \eq{suff1} hold. Then
 there exists $\eps_0>0$, $\gap>0$ and $\gam<-\ga$
  such that for all $\eps\in (0, \eps_0)$,
estimate \eq{rbound} holds with constants $C_0, C_1$ independent
of $\eps\in (0, \eps_0)$, and $M=M_0\eps^{-1}$, where $M_0$ is
also independent of $\eps$.
\end{lem}
\begin{proof}
 In view of \eq{chibound}, the
first three terms on the RHS of \eq{rphi} are bounded, and \[
\sup|\eps\chi'_\eps(x_2)|\to 0, \quad \eps\to 0, \]
 therefore it
suffices to show that $\gap>0$ and $\gam<-\ga$ can be chosen so
that
\begin{enumerate}[1)]
\item
$d_1+\ka_{11}^2/2+\ka_{21}\chi_\eps(x_2)-\be\chi_\eps(x_2)^2/2$ is
positive and bounded away from zero, uniformly in $x_2$ and
$\eps>0$;
\item
$d_2+\ka_{22}\chi_\eps(x_2)$ is positive and bounded away from 0,
uniformly in $x_2>0$ and $\eps>0$;
\item
$d_2+\ka_{22}\chi_\eps(x_2)$ is negative and bounded away from 0,
uniformly in $x_2<-1/\eps$ and $\eps>0$.
\end{enumerate}
Now, due to \eq{chim}, 3) is $d_2+\ka_{22}\gam<0$, which is
equivalent to \eq{gam}, and due to \eq{chip}, 2) reduces to
$d_2+\ka_{22}\gap>0$, which is trivial. On the strength of
\eq{chibound}, 1) is equivalent to the condition: the polynomial
\[
y\mapsto Q(y):=d_1+\frac{\ka_{11}^2}{2}+\ka_{21}y-\frac{\be}{2}y^2
\]
is positive on $[\gam, \gap]$. Since $\be>0$, it suffices to
choose $\ga_\pm$ so  that $Q(\ga_\pm)>0$. If $\gap>0$ is
sufficiently small, then $Q(\gap)$ is positive since
$Q(0)=d_1+\ka_{11}^2/2>0$ is, and if $Q(y)$ is positive at
$y=-\ga$ (which is condition \eq{suff1}), then it is positive for
$\gam$ in a sufficiently small  neighborhood of $-\ga$.
\end{proof}
\subsection{Proof of \theor{thm:reduction1}, part II: verification of conditions
(iv)--(v) of the general scheme}\label{part2}  When $\chi$ in the
definition of $\phi_\eps$ is fixed, the $L_\phi$ depends on $\eps$
only, and to stress the dependence on $\eps$, in this subsection,
we write $L_\eps$ instead of $L_\phi$. In the proofs of the
existence and the uniqueness theorem for the solution of problem
\eq{fc3}--\eq{bc3} and estimates \eq{resolvbound} and \eq{estinv},
we use an additional property of $r^0$, which was omitted from the
general scheme in \subsect{genscheme} as too technical (and by no
means necessary), and which holds for $r_{\infty, \phi}$ and
$r_{N, \phi}$. Fix  $\om\in (0, 1/2)$. We denote by $r_\eps$ any
function (in fact, a family of functions, parametrized by $\eps\in
(0, 1)$), which admits bound \eq{rbound} with  the constants
$C_1>0, C_2$ and $M=M_0\eps^{-1}$, where $C_0, C_1, M_0$ are
independent of $\eps\in (0, 1)$; in addition, $r_\eps$ must
satisfy the following Lipschitz condition with weight:
 \vskip0.1cm \noindent
there exists a constant $M_1>0$ independent of $\eps\in (0, 1)$
and such that for any $x, y\in \R_+\times\R$ satisfying
$|x_j-y_j|\le (|x_j|+\eps^{-1})^{\om}$, $j=1,2$,
\begin{equation}\label{rlipschitz}
|r_\eps(x)-r_\eps(y)|\le M_1(|x|+\eps^{-1})\eps^{1-\om}.
\end{equation}
 \begin{lem}\label{rcond} Functions $r_{\infty, \phi}$
and $r_{N, \phi}$ satisfy \eq{rbound} and \eq{rlipschitz},
uniformly in $N$ and $\eps\in (0, 1)$.
\end{lem}
\begin{proof} Each  of the functions $r_{\infty, \phi}$
and $r_{N, \phi}$ is the sum of a linear function $r$ or a
piece-wise linear function $r_N$, for which \eq{rlipschitz} is
evident since $\om\in (0, 1/2)$, and the function $\tr_\phi$ of
the form
\[
\tr_\phi(x)=b_{1, \eps}(x_2)x_1+ b_{2, \eps}(x_2)+ b_{0, \eps},
\]
where $b_{0, \eps}$ is bounded (hence, satisfies \eq{rlipschitz}),
and $b_{j, \eps}$, $j=1,2$, are constant outside the set $x_2\in
[-1/\eps, 0]$, and have derivatives of order $\eps$ on this set.
By applying the Lagrange theorem, we obtain \eq{rlipschitz} for
$b_{1, \eps}(x_2)x_1+ b_{2, \eps}(x_2)$. Thus, \eq{rlipschitz} has
been proved.

The proof of bound \eq{rbound} in \subsect{part1} is applicable to
any $r_\eps$, and the constants in \eq{rbound} can be chosen the
same for  all $\eps\in (0, 1)$.
\end{proof}
  The $-L_\eps$ is an elliptic operator in
the half-space $x_1>0$, which degenerates at the boundary $x_1=0$,
of the form
\begin{equation}\label{defL}
-L_\eps=-\theta_1\dd_1 -\frac{1}{2}x_1\dd_1^2
+\mu_\eps(x)\dd_2-\frac{\al+\be x_1}{2}\dd_2^2,
\end{equation}
where $\theta_1, \al, \be$ are positive, and $\mu_\eps$ is an
affine function of $x_1$
\begin{equation}\label{mucond}
\mu_\eps(x)=\mu_{\eps, 0}(x_2)+x_1\mu_{\eps, 1}(x_2),
\end{equation}
 whose coefficients are uniformly bounded, constant
 outside the segment $[-1/\eps, 0]$,
  and satisfy the global Lipschitz condition with parameter:
\begin{equation}\label{muest}
|\mu_{\eps, j}(y)-\mu_{\eps, j}(z)|\le C\eps|y-z|,
\end{equation}
where $C$ is independent of $\eps\in (0, 1)$.
\begin{thm}\label{thmest} Let $-L_\eps$ be an operator
of the form \eq{defL}, where $\theta_1, \al, \be$ are positive,
and the real-valued function $\mu_\eps$ satisfies \eq{mucond} and
\eq{muest}. Let $r_\eps$ satisfy \eq{rbound} and \eq{rlipschitz},
with  constants $C, C_0, C_1, M_0$  independent of $\eps\in (0,
1)$.

Then there exist $\eps_0>0$ such that for all $\eps\in (0,
\eps_0)$ \begin{enumerate}[a)] \item for any $T>0$ and any
continuous $g^0$, which exponentially decays at infinity, the
problem
\begin{eqnarray}\label{fc4}
(\dd_\tau-L_\eps+r_\eps)f(x, \tau)&=&0,\quad \tau\in (0,
T),\\\label{bc4} f(x, \tau)&=&g^0,
 \end{eqnarray} has the unique continuous
solution, which exponentially decays at infinity;
\item
set $\la_0=2M_0\eps^{-1}$, $\sg=\pi/2+\eps$; then there exists
$C_1$ and $\rho_0>0$ such that if $|\rho|\le \rho_0$  then for all
$\eps\in (0, \eps_0)$ and $\la\in \Sigma_{\sg, \la_0}$.
\begin{equation}\label{estinv2}
||\exp(\rho
\langle\cdot\rangle^{\prime\prime})(\la+r^0(\cdot))(\la-L_\eps+r_\eps)^{-1}\exp(-\rho
\langle\cdot\rangle^{\prime\prime})||\le C_1.
\end{equation}
\end{enumerate} \end{thm}
\begin{proof}
a) Instead of the problem on a strip $\tau\in [0, T]$, we can
consider the problem on $\tau\ge 0$, and look for the solution in
the class of continuous functions which admit a bound
\begin{equation}\label{ffbound} |f(x, \tau)|\le C\exp[-\de_1|x|+
\tau(\la_0-\de_1)],
\end{equation}
for some $\de_1>0$. By using the Laplace transform, we can reduce
the existence and uniqueness theorem to the uniform invertibility
 of the family $\la-L_\eps+r_\eps$, on the line $\Re\la=\la_0$;
the uniform invertibility follows from part b). Notice that in
order to prove that the solution satisfies the boundary condition
\eq{bc4} in the strong sense, the invertibility for $\la\in
\Sigma_{\sg, \la_0}$ and estimate \eq{resolvbound} are needed.

b) will be proved in the appendix. Notice that both
\eq{resolvbound} and \eq{estinv} are implied by \eq{estinv2}.
\end{proof}

\subsection{Proof of \theor{thm:reductionn}}\label{partn}
For $j=2,\ldots, n,$ fix
 $\ga_{+,j}>0$ and
 \begin{equation}\label{gamj}
 \ga_{-,j}<-\ga_j.
 \end{equation}
Next, take a non-decreasing function $\chi_j\in C^\infty(\R)$ such
that
\begin{equation}\label{chimj}
\chi_j(y)=\ga_{-,j},\quad y<-1;
\end{equation}
\begin{equation}\label{chipj}
\chi_j(y)=\ga_{+,j},\quad y>0;
\end{equation}
and
\begin{equation}\label{chiboundj}
\ga_{-,j}\le \chi_j(y)\le \ga_{+,j},\quad \forall\ y\in\R.
\end{equation}
Then, for any $\eps\in (0, 1)$, construct functions
$\chi_{j,\eps}(y)=\chi_j(\eps y)$ and
\[
\psi_{j,\eps}(y)=\int_0^y \chi_{j,\eps}(s)ds,
\]
and finally, set
\[
\phi_\eps(x)=\ka_{11} x_1 +\sum_{j=2}^n \psi_{j,\eps}(x_j).
\]
Clearly, it satisfies \eq{phibound}. Set \[\de:=\max\{\ka_{11},
\max_{j\ge 2}\ga_{+,j},
 \max_{j\ge 2}\{-\ga_j-\ga_{-,j}\}\};\]
  it is positive since $\ka_{11}>0,
\ga_{+,j}>0$ and \eq{gamj} holds.
\begin{lem}\label{l:boundf0n}
For any $\eps\in (0, 1)$, there exists $C_\eps$ such that for all
$x_1>0$ and  $x_2\in \R$,
\begin{equation}\label{f0bound1n} \exp(-\phi_{\eps}(x^{\prime\prime}))|f_0(r,
g; x, t)|\le C_\eps\exp(-\de(x_1+|x^{\prime\prime}|)).
\end{equation}
In particular, \eq{f0bound} holds. \end{lem} The proof of this
lemma is an evident modification of the proof of \lemm{l:boundf0n}
in the case $n=2$, and  the rest of the proof of
\theor{thm:reduction1} is a straightforward modification of
constructions and arguments in \subsect{part1} and
\subsect{part2}.

\subsection{Proof of \theor{thm:reduction23}}\label{reduction23}
Fix $\gam<-\ga$ and $\gap>0$, $l_1, l_2\in \R$, and for $\eps>0$,
construct $\chi_\eps$ and $\psi_\eps$  as in \sect{reductiona1n}.
After that, define
\[
\phi_\eps (x)=l_1x_1+l_2x_2+\psi_\eps(x_3).
\]
Set \[ \theta^1=\left[\begin{array}{l} \theta_1\\
\theta_2\end{array}\right], \quad
\dd^1=\left[\begin{array}{l} \dd_1\\
\dd_2\end{array}\right] \quad
x^1=\left[\begin{array}{l} x_1\\
x_2\end{array}\right]. \] We have
\begin{eqnarray*}
\exp(-\phi_\eps(x))L\exp(\phi_\eps(x))&=&
\langle\theta^1-\ka^{11}x^1, \dd^1+l^1\rangle\\
&&-(\ka_{31}x_1+\ka_{32}x_2+\ka_{33}x_3)(\dd_3+\chi_\eps(x_3))\\
&& +\frac{\be_{11}}{2}x_1 (\dd_1+l_1)^2+ \frac{\be_{22}}{2}x_2
(\dd_2+l_2)^2\\
&&
+\frac{1}{2}(\al_2+\be_{31}x_1+\be_{32}x_2)(\dd_3+\chi_\eps(x_3))^2,
\end{eqnarray*}
therefore
\begin{eqnarray*}
L_\phi&=& \langle \theta^1,
\dd^1\rangle+\sum_{j=1,2}\frac{\be_{jj}}{2}x_j\dd_j^2+\frac{1}{2}(\al_3+\be_{31}x_1+\be_{32}x_2)\dd_3^2\\
&& +((\be_{11}l_1-\ka_{11})x_1-\ka_{12}x_2)\dd_1+
((\be_{22}l_2-\ka_{22})x_2-\ka_{21}x_1)\dd_2\\
&&+[-\ka_{31}x_1-\ka_{32}x_2-\ka_{33}x_3+(\al_3+\be_{31}x_1+\be_{32}x_2)\chi_\eps]\dd_3,
\end{eqnarray*}
and by denoting the columns of the matrix $\ka^{11}$ as $\ka^j,
j=1,2,$
\begin{eqnarray*}
r_{\infty, \phi}(x)&=& d_0-\langle \theta^1, l^1\rangle
-\frac{\al_3}{2}(\eps\chi'_\eps(x_3)+\chi_\eps(x_3)^2)\\
&&+[d_1+\langle \ka^1,
l\rangle-\frac{\be_{11}l_1^2}{2}+\ka_{31}\chi_\eps(x_3)
-\frac{\be_{31}}{2}(\eps\chi'_\eps(x_3)+\chi_\eps(x_3)^2)]x_1\\
&&+[d_2+\langle \ka^2,
l\rangle-\frac{\be_{22}l_2^2}{2}+\ka_{32}\chi_\eps(x_3)
-\frac{\be_{32}}{2}(\eps\chi'_\eps(x_3)+\chi_\eps(x_3)^2)]x_2\\
&& +[d_3+\ka_{33}\chi_\eps(x_3)]x_3.
\end{eqnarray*}
Now it is clear what the optimal choice of $l_1$ and $l_2$ is.
Indeed, it is necessary that in the formula for $L_\phi$, the
coefficients at $\dd_j$, $j=1,2$, must be non-negative, hence
$l_j\ge \ka_{jj}/\be_{jj}$. On the other hand, in the formula for
$r_{\infty, \phi}$, it is better to have the coefficients at
$x_j$, $j=1,2$, as large as possible. Equivalently,
\[
\ka_{11}l_1-\frac{\be_{11}l_1^2}{2}+\ka_{21}l_2 \quad {\rm
and}\quad \ka_{22}l_2-\frac{\be_{22}l_2^2}{2}+\ka_{12}l_1
\]
should be as large as possible. But for $j=1,2$,
$\ka_{jj}l_j-\be_{jj}l_j^2/2$ attains its maximum at
$l_j=\ka_{jj}/\be_{jj}$, and since $\ka_{21}$ is negative,
$\ka_{21}l_2$ is attains its maximum on $[\ka_{22}/\be_{22},
+\infty)$  at $l_2=\ka_{22}/\be_{22}$. Hence, the optimal choice
is $l_j= \ka_{jj}/\be_{jj}, j=1,2,$ and with this choice, the
coefficients at $x_j, j=1,2,$ become
\[
d_1+\frac{\ka_{11}}{2\be_{11}}+\frac{\ka_{21}\ka_{22}}{\be_{22}}
+\ka_{31}\chi_\eps(x_3)
-\frac{\be_{31}}{2}(\eps\chi'_\eps(x_3)+\chi_\eps(x_3)^2)
\]
and
\[
d_2+\frac{\ka_{22}}{2\be_{22}}+\frac{\ka_{12}\ka_{11}}{\be_{11}}
+\ka_{32}\chi_\eps(x_3)
-\frac{\be_{32}}{2}(\eps\chi'_\eps(x_3)+\chi_\eps(x_3)^2).
\]
If $\eps>0$ and $\gap>0$ are sufficiently small, and $\gam<-\ga$
is sufficiently close to $-\ga$, then under conditions
\eq{suff21}--\eq{suff22}, both coefficients are positive and
bounded away from 0 uniformly in $x_3$. This property allows us to
repeat all the proofs in \sect{reductiona1n} with small and
evident changes.

There are two subtle point is in the construction of local
representatives and local almost inverses in the appendix: when
the set $U_{\eps, j}^2$ intersects with the plane $x_1=0$, we take
$x^{\eps, j}$ in this plane (similarly in the case of the
intersection with the plane $x_2=0$), and if with the line
$x_1=x_2=0$, then on this line. In the first case, we freeze all
the coefficients except for the ones in the expressions $x_1\dd_1$
and $x_1\dd_1^2$ (similarly for the local representatives at
points in the plane $x_2=0$), and in the second case, all the
coefficients except for the ones in the expressions $x_l\dd_l$ and
$x_l\dd_l^2$, $l=1,2$. In the first case, the model family are
operators on the half-line, and in the second case, on the
quadrant $\{x_1>0, x_2>0\}$. The remaining details are
straightforward modifications of the corresponding steps in the
proof for the family $A_1(2)$.

\appendix
\section{Proof of \theor{thmest}, b)}\label{proofs}
\subsection{Reduction to construction of almost inverses}
We start with the case $\rho=0$, and in the end of the proof,
indicate which changes need to be made for the case of small
$|\rho|$.

For each $\la\in \C$ satisfying $|\la|\ge \eps^{-1}$, introduce
the family of Banach (in fact, Hilbert) spaces
 $H^1_\la\subset H:=L_{2}(\R_+\times\R)$, which consists of
functions with the finite norm $||\cdot||_{H^1_\la}$ defined by
\[
||u||_{H^1_\la}^2:=||(|x|+|\la|)u||^2+||\dd_1 u||^2+ ||x_1 \dd_1
u||^2+\sum_{j=0}^2 (||\dd_2^j u||^2+||x_1\dd_2^j u||^2),
\]
where $||\cdot||$ is the norm in $H$.
Notice that the embedding $H^1_\la\subset H$ is continuous, and
$H^1_\la$ is independent of $\la$ as a topological space; but it
is convenient to derive estimates for the auxiliary operators
below in terms of the norm depending on the parameter.

To prove the invertibility of $-L_\eps+r_\eps+\la$ and estimate
\eq{resolvbound}, it suffices to construct left and right {\em
almost inverses}, $R_{\la, \eps}^l$ and $R_{\la, \eps}^r$. These
are operators which are uniformly bounded
\begin{eqnarray}\label{inverseboundl}
||R_{\la, \eps}^l||_{H\to H^1_\la}&\le& C;\\\label{inverseboundr}
||R_{\la, \eps}^r||_{H\to H^1_\la}&\le& C,
\end{eqnarray}
where $C$ is independent of $\eps\in (0, 1)$ and $\la\in
\Sigma_{\sg, \la_0}$, and satisfy the approximate equalities in
the definition of the inverse:
\begin{eqnarray}\label{inversel}
R_{\la, \eps}^l(-L_\eps+r_\eps+\la)&=&I+ T_{\la, \eps}^l,\\
\label{inverser} (-L_\eps+r_\eps+\la)R_{\la, \eps}^r&=&I+ T_{\la,
\eps}^r,
\end{eqnarray}
where
\begin{eqnarray}\label{inverseestr}
||T_{\la, \eps}^r||_{H\to H}&\to& 0, \quad {\rm as}\ \eps\to
0,\\\label{inverseestl} ||T_{\la, \eps}^l||_{H\to H^1_\la}&\to& 0,
\quad {\rm as}\ \eps\to 0,
\end{eqnarray}
uniformly w.r.t. $\la\in \Sigma_{\sg, \la_0}$. Notice that
\eq{inverser} and \eq{inverseestr} imply that $R_{\la, \eps}^r$
maps $H$ into the domain of $-L_\eps+r_\eps$. Suppose that the
almost inverses are constructed, and $\eps$ is so small that the
norms of operators $T_{\la, \eps}^r$ and $T_{\la, \eps}^l$ are
less than $1/2$. Then the RHS in \eq{inversel} and \eq{inverser}
are invertible bounded operators, and hence $-L_\eps+r_\eps+\la$
has the bounded inverse. Moreover, the inverse $R_{\la, \eps}^r
(I+T_{\la, \eps; r})^{-1}$ is a bounded operator from $H$ to
$H^1_\la$, uniformly in $\la$, which implies \eq{estinv2}.

Thus, it remains to construct left and right almost inverses. The
standard technique consists of the construction of
\begin{enumerate}[(i)]
\item
an appropriate partition of unity,
\item
 local representative of the operator (localization of the
operator), that is, the freezing of coefficients w.r.t. to all the
variables or some of them at some points,
\item
the inverses to the local representatives, and finally,
\item
global almost inverses by using the partition of unity and the
local inverses.
\end{enumerate}
Notice that $\Sigma_{\sg, \la_0}$ depends on $\eps$, and $|\la|\to
+\infty$ as $\eps\to 0$, uniformly in $\la\in \Sigma_{\sg,
\la_0}$; this property will be used systematically in the
constructions below.

\subsection{Partition of unity}
 We use the universal construction due to
H\"ormander (1985), which is based on the definition of a {\em
slowly varying metric}; for the variant of the construction for
operators with parameter, see Levendorski\v{i} (1993). Fix $\om\in
(0, 1/2)$, and for each $\eps\in (0, 1)$, define a function
$\langle y \rangle_\eps= (|\eps|^{-2}+y^2)^{1/2}$ on $\R$, and the
Riemann metric $G_\eps$ on on $R^2$ by
\[
G_{\eps; x}(z)=\langle x_1 \rangle_\eps^{-2\om}|z_1|^2+\langle x_2
\rangle_\eps^{-2\om}|z_2|^2.
\]
It is straightforward to check that the derivatives of the
function $\langle y \rangle_\eps^{\om}$ w.r.t. $y$ are bounded
uniformly in $y\in\R^2$ and  $\eps\in (0, 1)$, therefore (see
Levendorski\v{i} (1993)) there exist $c, c_1, C_1>0$ such that if
$G_{\eps; x}(x-z)\le c$, then
\begin{equation}\label{metrics}
c_1 G_{\eps; z}(w)\le G_{\eps; x}(w)\le C_1G_{\eps; z}(w),\quad
\forall\ w\in \R^2.
\end{equation}
Condition \eq{metrics} means that the metric $G_\eps$ is slowly
varying (uniformly w.r.t. $\eps\in (0, 1)$), and therefore there
exist positive constants $C_2, C_3, C_4$, points $x^{\eps, j}\in
\R^2,$ and non-negative functions $\nu_{\eps, j}\in
C^\infty(\R^2), j=1,2,\ldots,$ such that
\begin{enumerate}[(i)]
\item
sets $U_{\eps, j}=\supp \nu_{\eps, j}, j=1,2,\ldots,$ cover
$\R^2$;
\item
the multiplicity of the covering $\{U_{\eps, j}\}_{j\ge 1}$ is
bounded uniformly w.r.t. $\eps\in (0, 1)$,  and $j=1,2,\ldots,$
that is, for each $j$, the number of $k$ for which $U_{\eps, j}$
and $U_{\eps, k}$ intersect is bounded by $C_2$;
 \item $x^{\eps,
j}\in U_{\eps, j}$;
\item
the diameter of $U_{\eps, j}$ is not greater than
$C_2(|\eps|^{-1}+|x^{\eps, j}|)^{\om}$;
\item
for all multi-indices $s$, $\eps\in (0, 1)$,  and $j\ge 1$,
\begin{equation}\label{partition1}
|\dd^s \nu_{\eps, j}(x)|\le C_s\langle
x^{\eps,j}\rangle_\eps^{-|s|\om},\quad \forall\ x,
\end{equation}
where $|s|=s_1+s_2$, and the constants $C_s$ are independent of
$\eps\in (0, 1)$, and $j\ge 1$;
\item
for all $\eps\in (0, 1)$,  and all $x\in\R^2$,
\begin{equation}\label{partition2}
\sum_{j\ge 1}\nu_{\eps, j}(x)=1.
\end{equation}
\end{enumerate}
For details of the construction, see H\"ormander (1985) and
Levendorski\v{i} (1993). In the construction, one can choose the
points and functions so that

\vskip0.1cm \noindent (vii) either $x^{\eps, j}$ is on the line
$x_1=0$, or $x^{\eps, j}$ and $U_{\eps, j}$ are outside the strip
$|x_1|\le c_2 |\eps|^{-\om}$, where $c_2>0$ is independent of
$\eps\in (0, 1)$ and $j$.

\vskip0.1cm \noindent
 From now on, we assume that (i)--(vii) hold, and we consider only
 the points in the half-plane $x_1\ge 0$; now the $\nu_{\eps, j}$ are functions defined
 in the same half-plane, and \eq{partition2}
 holds for $x$ from this half-plane. We divide the set of the
 points $x^{\eps, j}$ in the half-plane $x_1\ge 0$
 into two subsets: $j\in J_0$, if $x^{\eps, j}$ is on the boundary
 $x_1=0$, and $j\in J_+$, if $x^{\eps, j}$ belongs to the open
 half-plane $x_1>0$.

 Construct
 \[
 \nu_{\eps, j}^1=\sum_{k: U_{\eps, k}\cap U_{\eps, j}\neq
 \emptyset}\nu_{\eps, k},
 \]
 and set $U^1_{\eps, j}=\supp U_{\eps, j}$. Similarly, starting with
 $\nu_{\eps, j}^1$
 and $U^1_{\eps, j}$, construct $\nu_{\eps, j}^2$
 and $U^2_{\eps, j}$
 Then $\nu_{\eps, j}^l$
 and $U^l_{\eps, j}$, $l=1,2,$ satisfy the same conditions as $\nu_{\eps, j}$
 and $U_{\eps, j}$ (with different constants) but \eq{partition2}.
 In addition,
 \begin{eqnarray}\label{nusupp1}
 \nu_{\eps, j}^1(x)&=&1,\quad \forall\ x\in U_{\eps, j},\\\label{nusupp2}
\nu_{\eps, j}^2(x)&=&1,\quad \forall\ x\in U_{\eps, j}^1.
\end{eqnarray}

\subsection{Local inverses to non-degenerate local representatives}
 For $j\in J_+$, denote by $A_{\la, \eps, j}$ the operator
 $-L_\eps+r_\eps+\la$ with the coefficients freezed at $x^{\eps,
 j}$:
\[
 A_{\la, \eps, j}= -\theta_1\dd_1-\frac{1}{2}x_1^{\eps, j}\dd_1^2+\mu_\eps(x^{\eps, j})\dd_2
 -\frac{\al+\be
 x_1^{\eps, j}}{2}\dd_2^2+ r_\eps(x^{\eps, j})+\la.\]
The function
\[
a_{\la, \eps, j}(\xi)=-i\theta_1\xi_1+\frac{1}{2}x_1^{\eps, j}\xi_1^2+i\mu_\eps(x^{\eps, j})\xi_2
+\frac{\al+\be
 x_1^{\eps, j}}{2}\xi_2^2+ r_\eps(x^{\eps, j})+\la\]
is the symbol of the operator $A_{\la, \eps, j}$, that is,
\[
A_{\la, \eps, j}=a_{\la, \eps, j}(D):=\cF^{-1}a_{\la, \eps,
j}(\xi)\cF,
\]
where $\cF$ is the Fourier transform. Equivalently, for a
sufficiently regular function $u$,
\[
a_{\la, \eps, j}(D)u(x)=(2\pi)^{-2}\int_{\R^2}e^{i\langle x,
\xi\rangle}a_{\la, \eps, j}(\xi)\hu(\xi)d\xi,
\]
where
\[
\hu(\xi)=\int_{\R^2}e^{-i\langle x, \xi\rangle}u(x)dx
\]
is the Fourier transform of $u$.
\begin{lem}
There exist $c>0$ and $\eps_0>0$ such that for all $\eps\in (0,
\eps_0)$, $\la\in\Sigma_{\sg, \la_0}$, $j\in J_+$ and $\xi\in
\R^2$,
\begin{equation}\label{invsymjplus}
|a_{\la, \eps, j}(\xi)|\ge c(|\la|+|x^{\eps, j}|+x_1^{\eps,
j}|\xi|^2).
\end{equation}
\end{lem}
\begin{proof}
Due to our choice of $\sg$,
\begin{equation}\label{lare}
\Re \la >-\eps_1 |\Im\la|,\quad \forall\ \la\in \Sigma_{\sg,
\la_0},
\end{equation}
where $\eps_1\to 0$ as $\eps\to 0$. Hence, if $C>0$ is fixed, then
for all $\la\in\Sigma_{\sg, \la_0}$, $j\in J_+$ and $\xi\in \R^2$,
satisfying
\begin{equation}\label{lasmall}
|\la|\le C(|x^{\eps, j}|+x_1^{\eps, j}|\xi|^2),
\end{equation}
we have
\[
\Re a_{\la, \eps, j}(\xi)\ge \frac{1}{2}x_1^{\la,
j}\xi_1^2+\frac{\al+\be
 x_1^{\eps, j}}{2}\xi_2^2+ r_\eps(x^{\eps, j})-C \eps_1(|x^{\eps, j}|+x_1^{\eps,
 j}|\xi|^2),
 \]
 and for sufficiently small $\eps>0$, \eq{invsymjplus} obtains.
 On the other hand, if
 \begin{equation}\label{lalarge}
|\la|\ge C(|x^{\eps, j}|+x_1^{\eps, j}|\xi|^2),
\end{equation}
then $|a_{\la, \eps, j}(\xi)-\la||\la|^{-1}\to 1$, as $
C\to+\infty,$ and hence  if $C$ is large, and $\la\in\Sigma_{\sg,
\la_0}$, $j\in J_+$ and $\xi\in \R^2$ satisfy condition
\eq{lalarge}, then \eq{invsymjplus} holds as well.
\end{proof}
In view of \eq{invsymjplus}, we can define  pseudo-differential
operators (PDO) $R_{\la, \eps, j}, j\in J_+,$ by
\[
R_{\la, \eps, j}=a_{\la, \eps, j}(D)^{-1}:=\cF^{-1}a_{\la, \eps,
j}(\xi)^{-1}\cF.
\]
By using the Fourier transform, we immediately obtain that
$R_{\la, \eps, j}$ is the inverse to $A_{\la, \eps, j}$:
\begin{equation}\label{invidentplus}
A_{\la, \eps, j}R_{\la, \eps, j}=I,\quad R_{\la, \eps, j}A_{\la,
\eps, j}=I.
\end{equation}
\subsection{Local inverses to degenerate local representatives}
For $j\in J_0$, define
 \[
 A_{\la, \eps,
 j}=-\theta_1\dd_1-\frac{1}{2}x_1\dd_1^2-\frac{\al+\be
 x_1}{2}\dd_2^2 +r_\eps(x^{\eps, j})+\la,
 \]
 and introduce the Hilbert space
 $H_{\la, 0}$ of functions on the half-plane $x_1>0$, with
 the finite norm $||\cdot||_{\la, \eps, j}$ defined by
 \[
 ||u||_{\la, \eps, j}^2=||\dd_1 u||^2+||x_1 \dd_1^2 u||^2
 +\sum_{s=0}^2(||\dd_2^s u||^2+||x_1\dd_2^s u||^2)+||(|x^{\eps, j}_2|+|\la|)u||^2,
 \]
 where $||\cdot||$ is the norm in $H=L_2(\R_+\times \R)$.
 \begin{lem}\label{l:inverse0}
 There exist $\eps_0>0$ and $C$ such that for each $\la\in\Sigma_{\sg, \la_0}$,
 the operator $A_{\la, \eps, j}: H_{\la, \eps, j}\to H$ is invertible, and
 the norm of the inverse, call it $R_{\la, \eps, j}$, is bounded by
 $C$.
 \end{lem}
 \begin{proof}
 This a very special case of general results for degenerate elliptic
 operators in Levendorski\v{i} (1993). We recall the scheme of the proof.
 First, we make the (partial) Fourier
transform $\tilde\cF$ w.r.t. to $x_2$,
 and obtain
 \[
A_{\la, \eps, j}=\tilde\cF^{-1}A_{\la, \eps, j}(\xi_2)\tilde\cF,
\]
where
\[
A_{\la, \eps,
j}(\xi_2)=-\theta_1\frac{d}{dy}-\frac{1}{2}y\frac{d^2}{dy^2}+
\frac{\al+\be y}{2}\xi_2^2+r_\eps(x^{\eps, j})+\la
\]
is the family of operators on $\R_+$ parametrized by $\xi_2\in\R$
and $\la\in\Sigma_{\sg, \la_0}$. Introduce the Hilbert space
 $H_{\la, \eps, j, \xi_2}$ of functions on $\R_+$, with
 the finite norm $||\cdot||_{\la, \eps, j, \xi_2}$ defined by
 \[
 ||u||_{\la, \eps, j, \xi_2}^2=||u'||^2+||y u^{\prime\prime}||^2+||(1+y)\xi^2
 u||^2+||(|x^{\eps, j}_2|+|\la|)u||^2,
 \]
 where $||\cdot||$ is the norm in $L_2(\R_+)$. By using an
appropriate partition on unity on $\R_+$, which depends on $(\la,
\xi_2)$, one shows that if $\eps_0$ is sufficiently small then for
all $\eps\in (0, \eps_0), \la\in\Sigma_{\sg, \la_0},$ and
$\xi_2\in\R$, operator
\[
A_{\la, \eps, j}(\xi_2): H_{\la, \eps, j, \xi_2}\to L_2(\R_+)
\]
is invertible, and the norm of the inverse, $R_{\la, \eps,
j}(\xi_2)$, is bounded by a constant which is independent of $\la,
\eps, \xi_2$.

It remains to define
\[
R_{\la, \eps, j}=\tilde\cF^{-1}R_{\la, \eps, j}(\xi_2)\tilde\cF.
\]
\end{proof}

\subsection{Construction of almost inverses}
Set \begin{eqnarray*}
 R_{\la, \eps}^l&=&\sum_{j\ge 1} \nu_{\eps, j}R_{\la,
\eps,
j}\nu^1_{\eps, j},\\
R_{\la, \eps}^r&=&\sum_{j\ge 1} \nu^1_{\eps, j}R_{\la, \eps,
j}\nu_{\eps, j}.
\end{eqnarray*}
We check that if $\eps_0>0$ is sufficiently small then the
conditions \eq{inverseboundr}, \eq{inverser} and \eq{inverseestr}
hold for the family of almost right inverses $R_{\la, \eps}^r$;
the verification of their analogs for almost left inverses
$R_{\la, \eps}^l$ is similar.

Notice that \begin{enumerate}[1)]
 \item
the multiplicity of the covering $\{U^2_{\eps, j}\}_{j\ge 1}$ is
finite;
\item
the norms are defined in terms of integration of a function and
its derivatives multiplied by weight functions;
\item the
derivatives of functions $\nu^s_{\eps, j}$ satisfy estimates
\eq{partition1}, and
\item \eq{partition2}, \eq{nusupp1} and
\eq{nusupp2} hold. \end{enumerate} Therefore, it suffices to prove
that
\begin{enumerate}[a)]
\item
there exist $\eps_0>0$ and $C$ such that for all $\eps\in (0,
\eps_0), \la\in\Sigma_{\sg, \la_0}$, and $j\ge 1$,
\begin{equation}\label{boundrloc}
||\nu^1_{\eps, j}R_{\la, \eps, j}||_{H\to H^1_\la}\le C;
\end{equation}
\item
let $\eps_0\to 0$; then uniformly in $\eps\in (0, \eps_0),
\la\in\Sigma_{\sg, \la_0}$, and $j\ge 1$,
\begin{equation}\label{localiz}
||\nu^2_{\eps, j}(-L_\eps+r_\eps+\la - A_{\la, \eps,
j})\nu^1_{\eps, j})||_{H^1_\la\to H}\to 0;,
\end{equation}
and
\begin{equation}\label{commute}
 ||\nu^2_{\eps, j}(\nu^1_{\la, j}A_{\la, \eps,
j} - A_{\la, \eps, j}\nu^1_{\eps, j})||_{H^1_\la\to H}\to 0.
\end{equation}
\end{enumerate}
Indeed, \eq{inverseboundr} follows from \eq{boundrloc} since the
multiplicity of the covering $\{U^1_{\eps, j}\}_{j\ge 1}$, call it
$M$, is finite: for any $u\in H$,
\begin{eqnarray*}
||R_{\la, \eps, r}u||_{H^1_\la}&\le &M \sup_{j}||\nu^1_{\eps,
j}R_{\la, \eps, j}\nu_{\eps, j}u||_H\\
&\le & MC\sup_{j}||\nu_{\eps, j}u||_H\\
&=&MC||u||_H.
\end{eqnarray*}
Further, by using \eq{nusupp2}, we can write
\begin{eqnarray}\label{ident}
(-L_\eps+r_\eps +\la)R_{\la, \eps, r}&=&\sum_{j\ge 1} \nu^1_{\eps,
j}A_{\la, \eps, j}R_{\la, \eps, j}\nu_{\eps, j}\\\nonumber
&&+\sum_{j\ge 1}\nu^2_{\eps, j}[\nu^1_{\la, j}A_{\la, \eps,
j}-A_{\la, \eps, j}\nu^1_{\la, j}] R_{\la, \eps, j}\nu_{\eps,
j}\\\nonumber
 &&+\sum_{j\ge 1}\nu^2_{\eps, j}(-L_\eps+r_\eps+\la -
A_{\la, \eps, j})\nu^1_{\eps, j})R_{\la, \eps, j}\nu_{\eps, j}.
\end{eqnarray}
By construction, $A_{\la, \eps, j}R_{\la, \eps, j}=I$, therefore
due to \eq{partition2} and \eq{nusupp1}, the first sum on the RHS
of \eq{ident} is identity:
\[\sum_{j\ge 1} \nu^1_{\eps,
j}A_{\la, \eps, j}R_{\la, \eps, j}\nu_{\eps, j}=\sum_{j\ge 1}
\nu^1_{\eps, j}\nu_{\eps, j}=I.
\]
Since the multiplicity of the covering $\{U^2_{\eps, j}\}_{j\ge
1}$ is finite, we conclude from \eq{boundrloc}, \eq{localiz} and
\eq{commute}, that the norms of the last two sums on the RHS of
\eq{ident}, as operators in $H$, tends to 0 as $\eps_0\to 0$,
uniformly in $\eps\in (0, \eps_0), \la\in\Sigma_{\sg, \la_0}$.
This proves \eq{inverser} and \eq{inverseestr}, and finishes the
proof of part b) of \theor{thmest}.

\subsection{Proof of estimates \eq{boundrloc}, \eq{localiz} and
\eq{commute}} It suffices to use the following simple
observations:

1. Due to the choice $\om\in (0, 1/2)$, for $j\in J_0$,
\eq{rlipschitz} gives
\[
|| (r_\eps (x)-r_\eps(x^{\eps, j}))\nu^1_{\eps, j}||_{H^1_\la\to
H}\to 0
\]
as $\eps\to 0$, uniformly in $j$;

2. Since function $\mu_\eps$ is an affine function of $x_1$ with
coefficients uniformly bounded w.r.t. $x_2$, $\eps\in (0, 1)$ and
the other parameters, and the coefficients satisfy \eq{muest}, we
conclude that
\[
 || \mu_\eps \nu^1_{\eps, j}\dd_2||_{H^1_\la\to H}\to 0
\]
as $\eps\to 0$, uniformly in $j$;

3. For $j\in J_+$, an error of freezing of any coefficient at
$x^{\eps, j}$, on the set $U^2_{\eps, j}$ produces a relative
error of order $\eps^\om$;

4. Any derivative of any $\nu_{\eps, j}$ and $\nu_{\eps, j}^l$ is
of order $\eps^\om$, and hence each time we calculate such a
derivative, the term, in which the derivative enters, has a small
norm.

\subsection{Proof of \eq{estinv} for  small $\rho$}
Set $L_{\eps, \rho}=\exp(\rho\langle
x_2\rangle)L_\eps\exp(-\rho\langle x_2\rangle)$. Direct
calculations show that $L_{\eps, \rho}\to L_\eps$ as $\rho\to 0$,
in the operator norm, uniformly in $\eps$, therefore for small
$\rho$, the almost inverses to $\la-L_\eps+r_\eps$ can be used to
construct the inverses to $\la-L_{\eps, \rho}+r_\eps$, and the
rest of the proof of \eq{estinv} remains the same.

\end{document}